\pdfoutput=1
\documentclass{article}

\usepackage[preprint, nonatbib]{neurips_2025}

\usepackage{subcaption}
\usepackage[utf8]{inputenc} 
\usepackage[T1]{fontenc}    
\usepackage{hyperref}       
\usepackage{url}            
\usepackage{booktabs}       
\usepackage{amsfonts}       
\usepackage{nicefrac}       
\usepackage{microtype}      
\usepackage[table]{xcolor}
\usepackage{lineno}
\usepackage{graphicx}
\usepackage{enumitem} 
\usepackage{float}
\usepackage{amsmath}
\usepackage{wrapfig}
\usepackage{caption}
\usepackage{enumerate}
\usepackage{amsthm}
\usepackage{multirow}
\usepackage{algorithm}
\usepackage{algorithmic}
\usepackage[normalem]{ulem}
\useunder{\uline}{\ul}{}

\title{DualEquiNet: A Dual-Space Hierarchical Equivariant Network for Large Biomolecules}

\author{
Junjie Xu$^1$, Jiahao Zhang$^1$, Mangal Prakash$^2$, Xiang Zhang$^1$, Suhang Wang$^1$ \\
$^1$The Pennsylvania State University \\
$^2$Independent Researcher \\ 
\texttt{\{junjiexu, jiahao.zhang, xzz89, szw494\}@psu.edu}, \texttt{mangalcrj@gmail.com}
}

\begin{document}

\maketitle

\begin{abstract}
    Geometric graph neural networks (GNNs) that respect E(3) symmetries have achieved strong performance on small molecule modeling, but they face scalability and expressiveness challenges when applied to large biomolecules such as RNA and proteins. These systems require models that can simultaneously capture fine-grained atomic interactions, long-range dependencies across spatially distant components, and biologically relevant hierarchical structure—such as atoms forming residues, which in turn form higher-order domains.
    Existing geometric GNNs, which typically operate exclusively in either Euclidean or Spherical Harmonics space, are limited in their ability to capture both the fine-scale atomic details and the long-range, symmetry-aware dependencies required for modeling the multi-scale structure of large biomolecules.
    We introduce DualEquiNet, a Dual-Space Hierarchical Equivariant Network that constructs complementary representations in both Euclidean and Spherical Harmonics spaces to capture local geometry and global symmetry-aware features.
    DualEquiNet employs bidirectional cross-space message passing and a novel Cross-Space Interaction Pooling mechanism to hierarchically aggregate atomic features into biologically meaningful units, such as residues, enabling efficient and expressive multi-scale modeling for large biomolecular systems.
    DualEquiNet achieves state-of-the-art performance on multiple existing benchmarks for RNA property prediction and protein modeling, and outperforms prior methods on two newly introduced 3D structural benchmarks demonstrating its broad effectiveness across a range of large biomolecule modeling tasks.
\end{abstract}

\section{Introduction}
\label{sec:intro}

Large biomolecules such as RNA and proteins play central roles in biological systems, enabling key functions ranging from enzymatic catalysis to molecular recognition and gene regulation~\cite{cate1996crystal, jumper2021highly, berman2000protein, sahin2014mrna, jiang2021rna, baker2001protein}. These molecules exist at the atomistic scale, where complex three-dimensional geometries, hierarchical organization (with atoms grouped into residues and domains), and long-range interactions jointly determine function. Crucially, molecular models must respect the underlying physical symmetries: predicted properties should transform consistently under global rotations and translations, remaining invariant for scalar quantities (such as energy) or equivariant for directional quantities (such as dipole moments). Designing computational models that can faithfully capture these geometric and symmetry-aware aspects is essential for accurate biomolecular modeling~\cite{han2022geometrically, townshend2021geometric, atz2021geometric}.

Despite recent advances in geometric deep learning, current geometric graph neural networks (GNNs) face two critical limitations when applied to large biomolecular systems. \textit{First}, most existing geometric GNNs model molecules purely at the atomic level, representing atoms as nodes and bonds or spatial proximities as edges~\cite{satorras2021n, duval2023hitchhiker, han2024survey}. While effective for small molecules, this atomistic-only approach fails to capture the inherent hierarchical organization in large biomolecules, where atoms are naturally grouped into higher-order units such as nucleotides (in RNA) or amino acids (in proteins). Ignoring this hierarchical structure leads to the loss of essential contextual and relational information, ultimately constraining predictive performance.
\textit{Second}, large biomolecules exhibit rich long-range dependencies: distant residues or nucleotides can interact and cooperatively determine global properties~\cite{deng2023rna}. Most existing geometric GNNs focus on local neighborhoods in Euclidean space~\cite{satorras2021n, schutt2017schnet, brandstetter2021geometric, cenhigh, aykent2025gotennet}, limiting their ability to capture nonlocal interactions effectively. While some methods incorporate higher-order geometric features, such as using higher-degree spherical harmonics, they often suffer from significant computational overhead~\cite{thomas2018tensor, brandstetter_geometric_2022, fuchs2020se}, making them impractical for large-scale molecular systems.

To address the challenges of hierarchical structure and long-range dependencies in large biomolecules, we propose DualEquiNet, a novel Dual-Space Hierarchical Equivariant Network. Unlike existing methods which either use nonlocal attention using spherical harmonics~\cite{frank2024euclidean} in Euclidean-space neighborhoods or using high-degree steerable representations within a single geometric space~\cite{cen2024high, aykent2025gotennet}, DualEquiNet jointly constructs complementary representations in both Euclidean (EU) and spherical harmonics (SH) spaces: the EU space captures local geometric interactions crucial for atomic-level detail, while the SH space encodes rotation-equivariant features well-suited for long-range structural dependencies. Through bidirectional cross-space message passing, DualEquiNet enables structured information exchange between EU and SH spaces, integrating local and global geometric contexts. Moreover, we introduce a novel Cross-Space Interaction Pooling (CSIP) mechanism that hierarchically aggregates atomic-level features into coarser scales (such as residue-level), expanding the receptive field without requiring deep stacking of atomic layers. By explicitly facilitating interaction between EU and SH representations during pooling, CSIP ensures efficient and expressive multi-scale modeling. Together, these innovations yield a scalable, symmetry-aware framework tailored to the unique architectural and functional demands of large biomolecular systems.

We evaluate DualEquiNet across established RNA and protein property prediction datasets, demonstrating consistent improvements over existing geometric baselines (with average performance gains of 8.4\%–33.5\% across tasks). To further probe the model’s ability to capture nuanced 3D structural features, we developed two new benchmark datasets for Solvent-Accessible Surface Area (SASA) Prediction and Torsion Angle Prediction, which are crucial for applications in drug discovery and protein engineering~\cite{shrake1973environment, stanton2022accelerating}. DualEquiNet achieves improvements of 3.1\%-28.8\% and 2.5\%, respectively, over the strongest prior methods on these benchmarks as well. 
To sum up, our \textbf{main contributions} are: 
\begin{itemize}[leftmargin=*]
\item (i) We introduce DualEquiNet, a novel dual-space hierarchical equivariant network that explicitly constructs complementary Euclidean and spherical harmonics (SH) representations, enabling the simultaneous capture of local atomic interactions and long-range structural dependencies. 
\item (ii) We propose bidirectional cross-space message passing and a novel Cross-Space Interaction Pooling (CSIP) mechanism, which together allow efficient and expressive hierarchical aggregation from atoms to residues, tailored to large biomolecular structures. 
\item (iii) We introduce two new 3D structural benchmark datasets (SASA and Torsion Angle prediction) and evaluate DualEquiNet across these and other diverse established biomolecular tasks, including RNA and protein property prediction, consistently achieving state-of-the-art results and outperforming prior baselines.
\end{itemize}

\section{Related Works} 
\label{sec:related_work}
\noindent\textbf{Invariant GNNs.}
Rotation-invariant GNNs leverage geometric features such as interatomic distances and angles to produce representations invariant under rigid body motions. SchNet~\cite{schutt2017schnet} uses continuous-filter convolutions over pairwise distances. DimeNet~\cite{gasteiger2020directional} and GemNet~\cite{gasteiger2021gemnet} further incorporate angular information, improving directional sensitivity. SphereNet~\cite{liu2022spherical} expands the invariant feature set to include distances, angles, and torsions. While effective, such models discard essential orientation-dependent information, motivating the need for equivariant approaches.

\noindent\textbf{Scalarization Equivariant GNNs.}
Scalarization-based equivariant GNNs aim to preserve E(3) equivariance by decoupling scalar and vector representations. EGNN~\cite{satorras2021n} uses pairwise distances to compute scalar messages for equivariant updates of coordinates. PaiNN~\cite{schutt2021equivariant} and GVPGNN~\cite{jing2020learning} similarly apply scalar gating and vector decoupling to maintain equivariance efficiently. However, their reliance on scalar guidance can limit expressivity, particularly for modeling complex rotational dependencies~\cite{joshi2023expressive}.

\noindent\textbf{Higher-order Equivariant GNNs.}
Tensor-product-based methods achieve full SE(3) or SO(3) equivariance via Clebsch--Gordan (CG) tensor algebra. TFN~\cite{thomas2018tensor} and SE(3)-Transformer~\cite{fuchs2020se} construct equivariant convolutions over irreducible representations, enabling expressive geometric reasoning. MACE~\cite{batatia2022mace} utilizes a complete basis from the Atomic Cluster Expansion for maximal expressivity. However, these methods often suffer from high computational cost due to CG operations scaling as $\mathcal{O}(L^6)$ with the degree $L$.
Recent methods aim to balance expressivity and efficiency by avoiding explicit CG products. HEGNN~\cite{cenhigh} and GotenNet~\cite{aykent2025gotennet} incorporate high-degree spherical harmonics but use inner-product-based scalarization to model cross-degree interactions.

\noindent\textbf{Long-range Dependencies.}
Modeling long-range interactions remains a core challenge in GNNs due to the tension between limited receptive fields and over-smoothing~\cite{chen2020simple, nguyen2023revisiting}. Recent benchmarks~\cite{dwivedi2022long, tonshoff2024where, zhou2025glora} evaluate GNN performance on long-range tasks. Solutions include implicit models (e.g., IGNNs) and Graph Transformers, which relax locality constraints via attention-based architectures. These efforts are complementary to equivariant modeling and may be integrated with geometric priors for improved scalability and fidelity.

A more comprehensive discussion of related work, including technical distinctions from our work, is provided in Appendix~\ref{appen:related}.

\section{Methodology} 
\label{sec:methodology}

\paragraph{Notations.} Each molecule can be represented as a 3D geometric graph, $\mathcal{G} = \{H, X \}$. $H$ contains node features across hierarchical scales, specifically atom-level (e.g. atomic number) and residue-level (e.g. nucleotide type in RNA), denoted by $H = \{ {H_{atom} \in \mathbb{R}^{N_{atom} \times F_{atom}}, H_{res} \in \mathbb{R}^{N_{res} \times F_{res}}}\}$, where $N_{atom}$ and $N_{res}$ are the number of atoms and residues (e.g. the nucleotides in RNA and amino acids in proteins), $F_{atom}$ and $F_{res}$ are the dimensions of atom and residue features. $h_{atom} \in \mathbb{R}^{F_{atom}}$ and $h_{res} \in \mathbb{R}^{F_{res}}$ are the corresponding features of a node.
$X \in \mathbb{R}^{N_{atom} \times 3}$ denotes the 3D coordinates of atoms. From initial $\mathcal{G}$, we aim to construct Euclidean (EU) space neighbors $\mathcal{N}_{EU}$ and spherical harmonics (SH) space neighbors $\mathcal{N}_{SH}$. 
We use uppercase letters to denote matrices and lowercase symbols to represent the corresponding vectors for a given node (e.g. $h_i$ and $\boldsymbol{x}_i$ for node $i$). To distinguish between invariant and equivariant features, we use regular letters for invariant features (e.g., $h$) and bold symbols for equivariant features (e.g., $\boldsymbol{x}$).

\begin{figure}[t]
    \begin{picture}(0,155)
    \put(0,0){\begin{minipage}[b]{0.285\textwidth}
    \caption*{(a) DualEquiNet   \ \ \ \ \ \ \  \ \ \ \ }
    \vspace{-0.5em}
    \includegraphics[width=\linewidth]{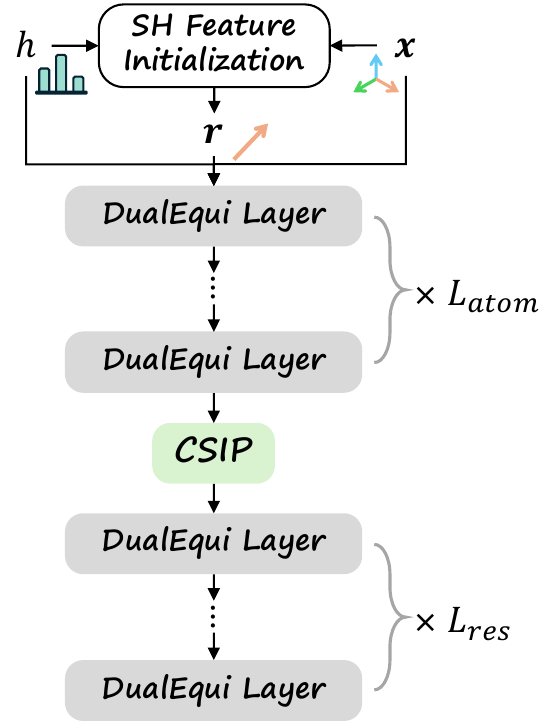}
    \end{minipage}}
        
    \put(113,0){\begin{minipage}[b]{.32\textwidth}
    \caption*{(b) DualEqui Layer}
    \vspace{-0.4em}
    \includegraphics[width=\linewidth]{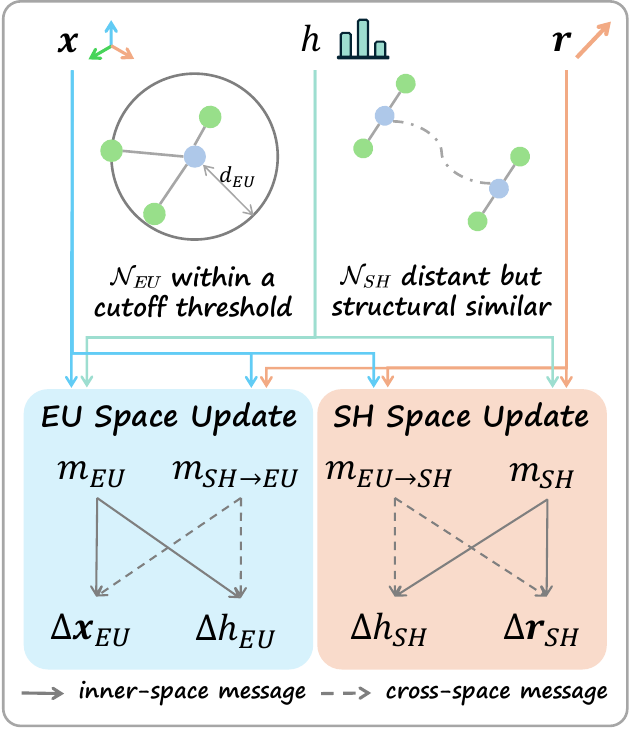}
    \end{minipage}}

    \put(245,49){\begin{minipage}[b]{.37\textwidth}
    \caption*{(c) CSIP}
    \vspace{-0.5em}
    \includegraphics[width=\linewidth]{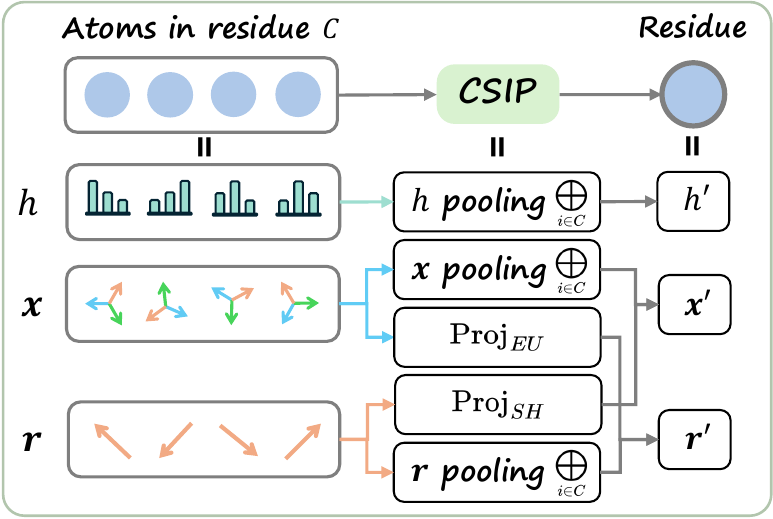}
    \end{minipage}}

    \put(260,-5){\begin{minipage}[b]{.3\textwidth}
    \caption*{\ \ \ \ \ \ \  \ \ \ \ (d) Neighborhoods}
    \vspace{-0.8em}
    \includegraphics[width=\linewidth]{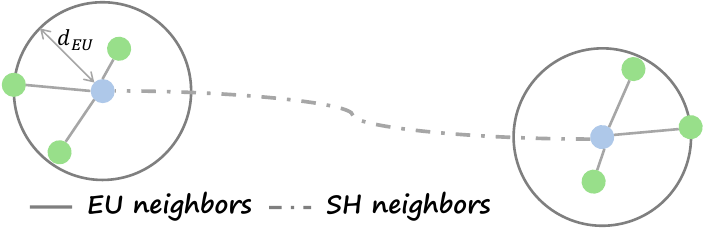}
    \end{minipage}}   
\end{picture}
\vspace{0.3em}
\caption{Overview of DualEquiNet. \textbf{(a) DualEquiNet}: It begins with SH feature initialization, followed by multiple DualEqui layers applied at the atomic level. A CSIP module then aggregates atomic representations into residue-level features, which are further refined by additional DualEqui layers. \textbf{(b) DualEqui Layer}: Each layer first infers EU and SH neighborhoods, then performs both inner-space and cross-space message passing to update the Euclidean coordinates $\boldsymbol{x}$, scalar features $h$, and SH features $\boldsymbol{r}$. \textbf{(c) CSIP}: Cross-Space Interaction Pooling aggregates atom-level features within each residue cluster. It performs pooling for $h$, $\boldsymbol{x}$, and $\boldsymbol{r}$, and incorporates cross-space projections to exchange complementary information across spaces. \textbf{(d) Neighborhoods}: EU neighbors are determined based on a distance cutoff, capturing local geometry. SH neighbors are selected based on SH feature similarity, enabling long-range structural connections beyond spatial proximity.
}
\label{fig:teaser}
\end{figure}

\textbf{Overview of DualEquiNet.} Fig.~\ref{fig:teaser}(a) gives an illustration of the proposed DualEquiNet. It begins with SH feature initialization, followed by multiple DualEqui layers applied at the atomic level. A Cross-Space Interaction Pooling (CSIP) module then aggregates atomic representations into residue-level features, which are further refined by additional DualEqui layers. Each DualEqui layer first infers EU and SH neighborhoods, then performs both inner-space and cross-space message passing to update the EU coordinates $\boldsymbol{x}$, invariant features $h$, and SH features $\boldsymbol{r}$ (See Fig.~\ref{fig:teaser}(b)). EU neighbors are determined based on a distance cutoff, capturing local geometry. SH neighbors are selected based on SH feature similarity, enabling long-range structural connections beyond spatial proximity (See Fig.~\ref{fig:teaser}(d)). CSIP aggregates atom-level features within each residue cluster. It performs pooling for $h$,  $\boldsymbol{x}$, and $\boldsymbol{r}$, and incorporates cross-space projections to exchange complementary information across spaces (See Fig.~\ref{fig:teaser}(c)). Next, we give details of each component.

\subsection{Initialization of Dual-Space Features}

Most equivariant graph neural networks rely on EU distance between nodes alone to define edges, which struggles to capture long-range dependencies. This often requires increasing distance thresholds or adding more layers, both of which increase computational complexity. Moreover, EU-based message passing assumes that neighboring nodes are close in EU space and share similar features, but this assumption doesn’t hold for large biomolecules like proteins and RNA. Nodes may be distant in EU space yet share similar structural features, such as bond angles and torsion, which are better captured in SH space. Thus, relying solely on EU proximity may miss key structural similarities crucial for molecular representation.

To address these issues, we propose a dual-space initialization strategy that constructs two complementary spaces: EU space for capturing local geometric interactions and SH space for global structural relationships. The EU neighborhood of node $i$ is defined as:
\begin{equation}
\label{eq:eu_neighbors}
    \mathcal{N}_{EU}(i) = \{ j \ | \  \lVert \boldsymbol{x}_i  - \boldsymbol{x}_j\rVert \leq d_{EU} \}, 
\end{equation}
where $d_{EU}$ is the Euclidean distance threshold. 
Next, we initialize the SH space features by aggregating the EU neighbors as follows. For each SH degree \( l = 0, 1, \ldots, l_{\text{max}} \), the \( l \)-th order SH feature for node \( i \) is defined as:
\begin{equation} 
\label{eq:init_sh_neigh}
    \boldsymbol{r}_{i}^{(l)} = \frac{1}{|\mathcal{N}_{EU}(i)|} \sum_{j \in \mathcal{N}_{EU}(i)} \phi_0 \left( \left[h_i, h_j, \left\|\boldsymbol{x}_{ij} \right\| \right] \right) \  \hat{Y}^{(l)} \left( \boldsymbol{\hat{x}}_{ij} \right),
\end{equation}
where $\boldsymbol{x}_{ij} = \boldsymbol{x}_{i} - \boldsymbol{x}_{j}$ is the direction vector, and $\boldsymbol{\hat{x}}_{ij} = \boldsymbol{x}_{ij} / \left\|\boldsymbol{x}_{ij} \right\|$ is the unit direction vector, $\phi_0$ is a differentiable message function (e.g., an MLP), and \( \hat{Y}^{(l)} \in \mathbb{R}^{2l + 1} \) is the normalized SH function of degree \( l \), with unit norm \( \| \hat{Y}^{(l)}(\cdot) \| = 1 \). 
The SH basis and corresponding node features up to degree \( l_{\text{max}} \) are jointly represented as: 
\begin{equation}\label{eq:sh_feat_all_order}
    \hat{Y} \left( \boldsymbol{\hat{x}}_{ij} \right) = [\hat{Y}^{(0)} \left( \boldsymbol{\hat{x}}_{ij} \right), \hat{Y}^{(1)} \left( \boldsymbol{\hat{x}}_{ij} \right), \cdots, \hat{Y}^{(l_{max})} \left( \boldsymbol{\hat{x}}_{ij} \right) ] , 
    ~~ \text{and} ~~
    \boldsymbol{r}_i = [\boldsymbol{r}_i^{(0)}, \boldsymbol{r}_i^{(1)}, \cdots, \boldsymbol{r}_i^{(l_{max})}] . 
\end{equation}
The details of spherical harmonics are introduced in Appendix~\ref{appen:math}.
Eq.~\ref{eq:init_sh_neigh} aims to learn an initialization of the SH space features based on the EU neighbors of the node, after which the SH features are further refined through updates in the SH space. 
With the initial SH vectors, we define the SH neighborhoods of node $i$ based on the cosine similarity as 
\begin{equation} 
    \mathcal{N}_{SH}(i) = \Big\{  j \ \Big| \  \frac{ \boldsymbol{r}_{i} \cdot \boldsymbol{r}_{j}}{\|\boldsymbol{r}_{i}\| \cdot \|\boldsymbol{r}_{j}\| } \geq d_{SH} \Big\} , 
\end{equation}
where $d_{SH}$ is a threshold for SH space. These two neighborhoods—EU and SH—form the basis for equivariant dual-space message passing and cross-space interaction in subsequent model stages. In this way, the SH neighborhood overcomes the distance limitations of EU space and captures relationships between nodes that are distant in EU space.

\subsection{DualEqui Layer}
The DualEqui Layer integrates geometric information from both EU and SH spaces through a bidirectional message passing scheme. Each layer updates invariant features $h$, EU coordinates $\boldsymbol{x}$, and SH features $\boldsymbol{r}$ using a combination of within-space and cross-space messages. This design enables the model to simultaneously preserve local geometry and capture global structural patterns.

\noindent\textbf{Euclidean Space Update}. In EU space, both the atomic positions $\boldsymbol{x}$ and invariant features $h$ are updated using information from two sources: (1) within-space interactions based on EU neighbors, and (2) cross-space interactions from the SH neighbors. Specifically, we compute messages from both the EU and SH neighbors and aggregate them via a multi-head attention mechanism. 

Within-space message $m_{EU}$ captures local geometric interactions in the EU space using invariant features and distances, while cross-space message $m_{SH \to EU}$ models the influence of SH-space structural similarity on the EU update, enabling long-range information flow guided by symmetry-aware features. For each node $i$ and its neighbor $j$, we have 
\begin{equation}
\label{eq:eu_space_msg}
    m_{EU, ij} = \phi_{EU} \left( \left[h_i, h_j, \left\|\boldsymbol{x}_{ij} \right\| \right] \right) , \quad
    m_{SH \to EU, ij} = \psi_{SH \to EU} ([\boldsymbol{r}_i \odot \boldsymbol{r}_j, \left\|\boldsymbol{x}_{ij} \right\|]), 
\end{equation}
where $\| \boldsymbol{x}_{ij} \|$ denotes the Euclidean distance between nodes $i$ and $j$. The function $\phi_{\text{EU}}$ denotes differentiable functions applied within EU space (e.g., an MLP), and $\psi_{SH \to EU}$ denotes cross-space transformations. The operator $\odot$ computes degree-wise inner products of SH features to form an invariant representation~\cite{cen2024high}:
\begin{equation}
\label{eq:sh_space_odot} 
    \boldsymbol{r}_i \odot \boldsymbol{r}_j = \left[\boldsymbol{r}_i^{(0)} \cdot \boldsymbol{r}_j^{(0)}, \boldsymbol{r}_i^{(1)} \cdot \boldsymbol{r}_j^{(1)}, \dots, \boldsymbol{r}_i^{(l_{\text{max}})} \cdot \boldsymbol{r}_j^{(l_{\text{max}})}\right].
\end{equation}
To combine the messages, we use $K$-head attention. Each head learns to attend to different aspects of the neighborhood, both in the EU and SH spaces. For the $k$-th head, the attention weights are computed as: 
\begin{equation}
    \alpha_{EU, ij}^k = \sigma\left( \phi_{EU, att}^k (m_{EU, ij}) \right) , 
    \quad 
    \beta_{EU, ij}^k = \sigma\left( \psi_{EU, \beta}^k (m_{SH \to EU, ij}) \right) , 
\end{equation}
where $\sigma$ is the sigmoid function, and $\phi_{\text{EU}, \text{att}}^k$, $\psi_{\text{EU}, \beta}^k$ are attention blocks.
The invariant feature $h_i$ is updated using a permutation-invariant aggregation $\bigoplus$ (e.g., sum or mean) over messages from both sets of neighbors:
\begin{equation} 
\label{eq:delta_h_e}
    \Delta h_{EU, i} = \frac{1}{K} \sum_{k=1}^K \phi_{upd, EU} \Big( \Big[ h_i, \bigoplus_{j \in \mathcal{N}_{EU}(i)} \alpha_{EU, ij}^k \ \phi_{EU, h}^k (m_{EU, ij}) + \bigoplus_{j \in \mathcal{N}_{SH}(i)} \beta_{EU, ij}^k \psi_{EU, h}^k (m_{SH \to EU, ij}) \Big] \Big),
\end{equation}
where, $\phi_{EU, h}^k$ and $\psi_{EU, h}^k$ are learnable message transformation networks per head.
Similarly, the coordinate update is computed by aggregating directional messages scaled by attention:
\begin{equation} 
\label{eq:delta_x_e}
    \Delta \boldsymbol{x}_{EU, i} = \frac{1}{K} \sum_{k=1}^K \Big( \bigoplus_{j \in \mathcal{N}_{EU}(i)} \alpha_{EU, ij}^k \ \phi_{EU, x} \left( m_{EU, ij} \right) \boldsymbol{x}_{ij} + \bigoplus_{j \in \mathcal{N}_{SH}(i)} \beta_{EU, ij}^k \psi_{EU, x}(m_{SH \to EU, ij}) \boldsymbol{x}_{ij} \Big)
\end{equation}
The EU space update integrates both local geometry and global SH information through learned, attention-weighted message passing. This allows the model to effectively capture rich interactions that span different geometric representations.

\noindent\textbf{Spherical Harmonics Space Update} Similarly, in SH space, for each node $i$ and its neighbor $j$, we compute messages from both the EU and SH neighbors as
\begin{equation}
\label{eq:sh_space_msg}
    m_{SH, ij} = \phi_{SH} \left( \left[h_i, h_j, r_i \odot r_j \right] \right) ,  \quad 
    m_{EU \to SH, ij} = \psi_{EU \to SH} ([\boldsymbol{r}_i \odot \boldsymbol{r}_j, \left\|\boldsymbol{x}_{ij} \right\|])
\end{equation}
We then adopt $K$-head attention to aggregate the information as
\begin{equation} 
\label{eq:delta_h_sh}
    \Delta h_{SH, i} = \frac{1}{K} \sum_{k=1}^K \phi_{upd, SH} \Big( \Big[ h_i, \bigoplus_{j \in \mathcal{N}_{SH}(i)} \alpha_{SH, ij}^k \ \phi_{SH, h}^k (m_{SH, ij}) + \bigoplus_{j \in \mathcal{N}_{EU}(i)} \beta_{SH, ij}^k \psi_{SH, h}^k (m_{EU \to SH, ij}) \Big] \Big) ,  
\end{equation}
where $\alpha_{SH, ij}^k = \sigma\left( \phi_{SH, att}^k (m_{SH, ij}) \right)$ and $
\beta_{SH, ij}^k = \sigma\left( \psi_{SH, \beta}^k (m_{EU \to SH, ij}) \right)$.
The SH feature is affected by $\boldsymbol{r}_j$ and $\hat{Y}(\boldsymbol{\hat{x}}_{ij})$ together. Specifically, for the within SH space update, utilizing $\boldsymbol{r}_j$ directly captures the structural similarity, including angle and torsion features, among SH neighbors. For cross-space updates from EU to SH, the relative direction of the EU neighbors is crucial. $\hat{Y}(\boldsymbol{\hat{x}}_{ij})$ explicitly incorporates this geometric directional information from its EU neighbors, potentially impacting SH features with the update given by the following. 
{
\begin{equation}
\label{eq:delta_r_sh}
    \Delta \boldsymbol{r}_{SH, i} = \frac{1}{K} \sum_{k=1}^K \Big( \bigoplus_{j \in \mathcal{N}_{SH}(i)} \alpha_{SH, ij}^k \phi_{SH, r} (m_{SH, ij}) \boldsymbol{r}_j + \bigoplus_{j \in \mathcal{N}_{EU}(i)} \beta_{SH, ij}^k \psi_{SH, r}(m_{EU \to SH, ij}) \hat{Y}(\boldsymbol{\hat{x}}_{ij}) \Big)
\end{equation}}
\noindent\textbf{Dual Space Update}. We employ a residual connection, and the overall update is given by
\begin{equation}
\label{eq:residual_update}
    h_i = h_i + \Delta h_{EU, i} + \Delta h_{SH, i}  ,  \quad
    \boldsymbol{x}_i = \boldsymbol{x}_i + \Delta \boldsymbol{x}_{EU, i}  ,  \quad
    \boldsymbol{r}_i = \boldsymbol{r}_i + \Delta \boldsymbol{r}_{SH, i}  .
\end{equation}




\subsection{Cross-space Interaction Pooling}
To create residue- and molecule-level representations, we use hierarchical pooling over atomic node clusters. The challenge is to maintain both local geometry and global structural context. Standard pooling methods, like average or attention-based aggregation, handle features from different geometric spaces separately. In biomolecular structures, local and global geometry are interconnected; for instance, an RNA nucleotide's torsional configuration (well-represented in SH space) can be affected by nearby atomic arrangements (in EU space), and vice versa.

To address the challenge, we introduce CSIP, a novel hierarchical pooling strategy that facilitates bidirectional information exchange between EU (local) and SH (global) spaces during aggregation. This approach integrates multi-scale geometric information while preserving equivariance.
Specifically, we first compute attention weights $\alpha_i$ using only invariant quantities, then the weights are used to perform attention-based pooling as:
\begin{equation} 
\label{eq:h_pooling}
    h' = \frac{1}{|C|} {\sum}_{i \in C} \alpha_i h_i , 
    \quad \quad
    \alpha_i = \sigma(\phi_{P}([h_i, d_i])) , 
\end{equation}
where $\phi_{P}$ is a neural network applied to invariant features to calculate the attention score, $d_i$ denotes the average pairwise distance from node $i$ to other nodes, and $C$ is the cluster of nodes over which pooling is performed. For example, a nucleotide in RNA or amino acid in protein consists of a cluster of atoms. 
To enrich the EU representation with global structural cues, we inject projected SH features: 
\begin{equation} 
\label{eq:x_pooling}
    \boldsymbol{x'} = \frac{1}{|C|} {\sum}_{i \in C} \alpha_i \left[ \boldsymbol{x}_i + \gamma \cdot \text{Proj}_{SH}(\boldsymbol{r}_i) \right], \quad
    \text{Proj}_{SH}(\boldsymbol{r}_i) = {\sum}_{l=1}^{l_{max}} w_l \| \boldsymbol{r}_i^{(l)} \| \boldsymbol{x}_i , 
\end{equation}
where \(  \boldsymbol{r}_i^{(l)} \) is the SH component at degree \( l \), and \( w_l \) are learnable weights. $\text{Proj}_{SH}$ converts SH space features $\boldsymbol{r}$ to EU space by producing an invariant signal summarizing the SH features' norms, which modulates the geometric position via a learnable coefficient $\gamma$.
Moreover, we enhance SH features with a projection of relative EU coordinates: 
\begin{equation} 
\label{eq:r_pooling}
    \boldsymbol{r}' = \frac{1}{|C|} {\sum}_{i \in C} \alpha_i \left[ \boldsymbol{r}_i + \epsilon \cdot \text{Proj}_{EU}(\boldsymbol{x}_i) \right], \quad
    \text{Proj}_{EU}(\boldsymbol{x}_i) = \hat{Y}(\boldsymbol{x}_i - \boldsymbol{x}_{\text{avg}}) , 
\end{equation}
where $\boldsymbol{x}_{\text{avg}} = \frac{1}{|C|} \sum_{i \in C} \boldsymbol{x}_i$ centers the input to maintain translation equivariance. The projection \( \text{Proj}_{EU}(x_i) \) converts $\boldsymbol{x}$ from EU space to SH space. Therefore, we choose to use the SH function $\hat{Y}$. It embeds local relative geometry into the SH domain, controlled by a learnable coefficient $\epsilon$. The pooled features $h'$, $\boldsymbol{x}'$, and $\boldsymbol{r}'$ serve as residue-level inputs for the subsequent DualEqui layers, enabling hierarchical modeling from atoms to residues.

\subsection{DualEquiNet}
As shown in Figure~\ref{fig:teaser}(a), DualEquiNet captures the hierarchical nature of biomolecular structures through two sequential modules: an initial atom-level processing module and a subsequent residue-level module, separated by the CSIP layer. Each module consists of multiple DualEqui layers. Figure~\ref{fig:teaser}(b) shows that each DualEqui layer encompasses the operations defined in Equations~\ref{eq:eu_neighbors} to~\ref{eq:residual_update}, and the network-level update is expressed as:
\begin{equation}
h^{l+1}, \boldsymbol{x}^{l+1}, \boldsymbol{r}^{l+1} = \text{DualEquiLayer}(h^{l}, \boldsymbol{x}^{l}, \boldsymbol{r}^{l}),
\end{equation}
where the superscript ${l}$ denotes the layer index, and is distinct from the SH order $(l)$ with parenthesis. There are two types of layers: atom layers and residue layers. The model first applies $L_{atom}$ DualEqui layers to capture fine-grained atomic interactions in both EU and SH spaces. The CSIP module then aggregates these features into residue-level representations, which are further processed by $L_{res}$ residue-level DualEqui layers to capture higher-order structural patterns and long-range dependencies.
The input to the first atom layer is $h^0 = h_{atom}$. The first residue layer receives $h^{L_{atom}+1} = [h^{L_{atom}}, h_{res}]$, a concatenation of the final atom-layer output and the CSIP-derived residue features. This hierarchical design enhances both computational efficiency and representational capacity, aligning naturally with the multi-scale structure of biomolecules.
The equivariance of the networks is proved in Appendix~\ref{appen:proofs}.

\section{Experiments and Results} 
\label{sec:exp}

We evaluate DualEquiNet across a suite of tasks to assess its expressivity, ability to capture long-range dependencies, and effectiveness on real-world biomolecular property prediction. Our experiments span synthetic benchmarks and challenging biomolecular tasks, including RNA property prediction, solvent-accessible surface area estimation, and torsion angle regression. The dataset details are in Appendix~\ref{appen:datasets_detials}. 

We benchmark DualEquiNet against baselines from various categories: (1)Invariant: SchNet~\cite{schutt2017schnet}; (2)Scalarization Equivariant: EGNN~\cite{satorras2021n}, GVPGNN~\cite{jing2020learning}; (3)Higher-order Equivariant: TFN~\cite{thomas2018tensor}, FastEGNN~\cite{zhangimproving}, HEGNN~\cite{cen2024high}, and GotenNet~\cite{aykent2025gotennet}, covering a range of popular and recent baselines including invariant, equivariant and low- to high-order tensor methods. A detailed baseline description is provided in Appendix~\ref{appen:compare_baselines}.

\subsection{Model Expressivity on Synthetic Dataset}
\label{sec:synthetic}

\begin{table}[t]
\caption{(a) Left: The $N$-chain task aims to differentiate two $N$-chain geometric graphs that differ only in the positions of their end nodes. (b) Right: The expressivity in the $N$-chain task with $N=10$. Anomalous results are highlighted in \colorbox{red!10}{red} and expected in \colorbox{green!10}{green}. DualEquiNet demonstrates a stronger capability to differentiate the two geometric graphs due to its expressive architecture.}
\label{tab:kchain}
    \centering
    \begin{minipage}{0.29\textwidth}
        \centering
        \includegraphics[width=\textwidth]{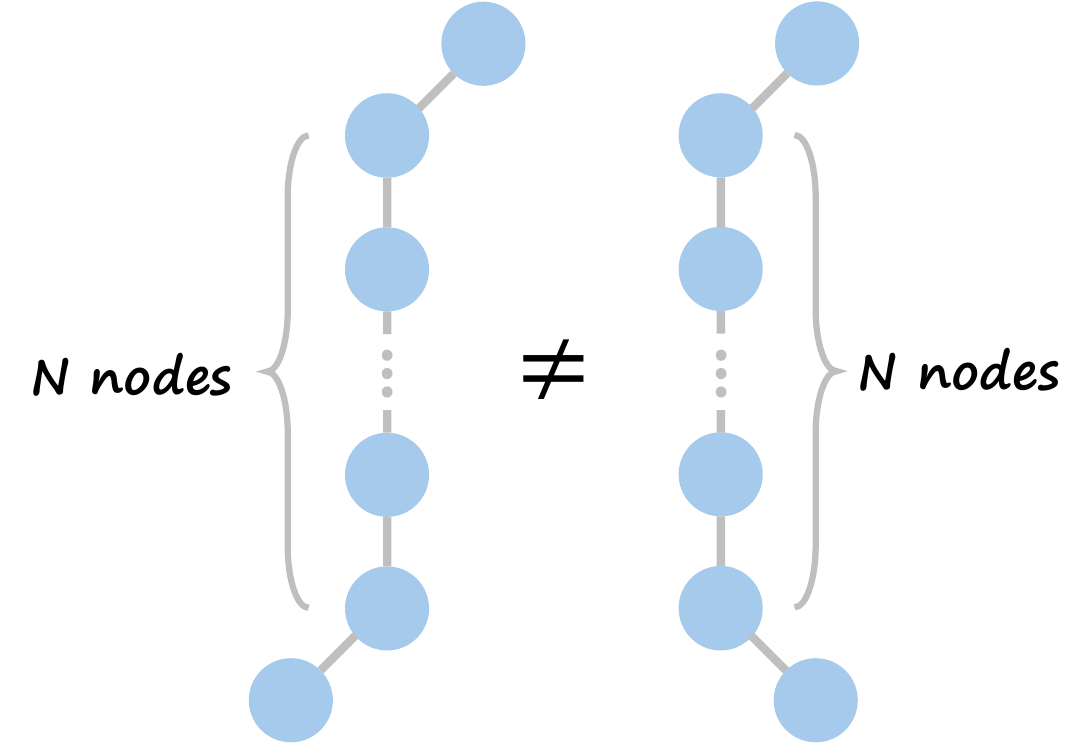}
    \end{minipage}
    \hfill
    \begin{minipage}{0.7\textwidth}
        \centering
        \label{tab:express_kchains}
        \resizebox{\textwidth}{!}{
        \begin{tabular}{ccccccccc}
        \toprule
        \textbf{Layers}         & $\lfloor \frac{N}{2} \rfloor = $ \textbf{5}          & $\lfloor \frac{N}{2} \rfloor + 1 =$ \textbf{6}           & \textbf{7}           & \textbf{8}           & \textbf{9}           & \textbf{10}         \\
        \midrule
        SchNet         & 50.0\footnotesize±0.0  & 50.0\footnotesize±0.0  & 50.0\footnotesize±0.0  & 50.0\footnotesize±0.0  & 50.0\footnotesize±0.0  & 50.0\footnotesize±0.0  \\
        EGNN           & 50.0\footnotesize±0.0  & 50.0\footnotesize±0.0  & 50.0\footnotesize±0.0  & 50.0\footnotesize±0.0  & 50.0\footnotesize±0.0  & 50.0\footnotesize±0.0  \\
        GVPGNN         & 50.0\footnotesize±0.0  & 52.0\footnotesize±9.8  & 51.0\footnotesize±7.0  & 53.0\footnotesize±11.9 & 51.5\footnotesize±8.5  & 54.5\footnotesize±14.3 \\
        TFN            & 50.0\footnotesize±0.0  & 50.0\footnotesize±0.0  & 50.0\footnotesize±0.0  & 50.0\footnotesize±0.0  & 50.0\footnotesize±0.0  & 50.0\footnotesize±0.0  \\
        FastEGNN       & 99.5\footnotesize±5.0  & 100.0\footnotesize±0.0 & 91.0\footnotesize±19.2 & 99.5\footnotesize±5.0  & 100.0\footnotesize±0.0 & 51.5\footnotesize±8.5  \\
        HEGNN          & 50.0\footnotesize±0.0  & 50.0\footnotesize±0.0  & 50.0\footnotesize±0.0  & 50.0\footnotesize±0.0  & 50.0\footnotesize±0.0  & 50.0\footnotesize±0.0  \\
        GotenNet       & 50.0\footnotesize±0.0  & 50.0\footnotesize±0.0  & 50.0\footnotesize±0.0  & 50.0\footnotesize±0.0  & 50.0\footnotesize±0.0  & 50.0\footnotesize±0.0  \\
        \midrule
        DualEquiNet $d_{SH}$=0.95  & 52.5\footnotesize±10.9 & 51.5\footnotesize±8.5  & 52.5\footnotesize±10.9 & 53.5\footnotesize±12.8 & \cellcolor{green!10} 99.5\footnotesize±5.0  & \cellcolor{red!10} 52.5\footnotesize±13.0 \\
        DualEquiNet $d_{SH}$=0.90  & 98.5\footnotesize±8.5  & 50.5\footnotesize±5.0  & 50.5\footnotesize±5.0  & \cellcolor{green!10} 99.0\footnotesize±7.0  & \cellcolor{green!10} 99.5\footnotesize±5.0  & \cellcolor{red!10} 53.5\footnotesize±12.8 \\
        DualEquiNet $d_{SH}$=0.80  & 98.0\footnotesize±9.8  & 54.0\footnotesize±13.6 & \cellcolor{green!10} 100.0\footnotesize±0.0 & \cellcolor{green!10} 100.0\footnotesize±0.0 & \cellcolor{green!10} 100.0\footnotesize±0.0 & \cellcolor{green!10} 100.0\footnotesize±0.0 \\
        DualEquiNet $d_{SH}$=0.70  & \cellcolor{green!10} 99.0\footnotesize±7.0  & \cellcolor{green!10} 100.0\footnotesize±0.0 & \cellcolor{green!10} 100.0\footnotesize±0.0 & \cellcolor{green!10} 100.0\footnotesize±0.0 & \cellcolor{green!10} 100.0\footnotesize±0.0 & \cellcolor{green!10} 100.0\footnotesize±0.0 \\
        DualEquiNet $d_{SH}$=0.60  & \cellcolor{green!10} 100.0\footnotesize±0.0 & \cellcolor{green!10} 100.0\footnotesize±0.0 & \cellcolor{green!10} 100.0\footnotesize±0.0 & \cellcolor{green!10} 100.0\footnotesize±0.0 & \cellcolor{green!10} 100.0\footnotesize±0.0 & \cellcolor{green!10} 100.0\footnotesize±0.0 \\
        \bottomrule
        \end{tabular}
        }
    \end{minipage}
\end{table}

To evaluate DualEquiNet’s ability to capture long-range dependencies, we use the $N$-chain discrimination task from prior work~\cite{joshi2023expressive}. This task provides a synthetic yet rigorous benchmark for assessing the expressive power of geometric GNNs, particularly their capacity to propagate and integrate information across distant nodes. In this experiment, as illustrated in Table~\ref{tab:kchain} (left), there are a pair of graphs and each graph consists of a linear chain with $N$ nodes, identical except for their end-node configurations—analogous to chiral molecules. The model is trained on one graph and evaluated on both to test its ability to distinguish the chiral pair. These graphs are ($\lfloor \frac{N}{2} \rfloor + 1$)-hop distinguishable, meaning a GNN requires at least that many layers to differentiate them as proven in~\cite{joshi2023expressive}. As $N$ increases, the task becomes more challenging due to oversquashing~\cite{alon2021on}. While the original Geometric Weisfeiler-Leman (GWL) test uses $N=4$, we extend this to $N=10$, requiring at least 6-hop information—mirroring challenges in modeling large biomolecules. Note that in real-world scenarios, large biomolecules can consist of thousands of nodes, making long-range information exchange even more challenging. We vary the SH-space cutoff distance $d_{SH}$ to study its effect on DualEquiNet performance. 

As shown in Table~\ref{tab:kchain}, most baselines fail to distinguish the $N$-chain graphs, performing no better than random guessing even when stacking more than $\lfloor \frac{N}{2} \rfloor + 1$ layers. FastEGNN is an exception due to its global virtual nodes. In contrast, DualEquiNet achieves higher accuracy even with fewer than $\lfloor \frac{N}{2} \rfloor + 1$ layers by leveraging SH-space neighbors for global context. Performance improves as $d_{SH}$ decreases, reaching near-perfect accuracy at $d_{SH} \leq 0.7$, demonstrating the effectiveness of SH-space communication in overcoming long-range limitations with lower SH cutoff distance signifying more SH-space neighbors. We also validate expressivity on another rotational symmetry task (Appendix~\ref{appen:exp_rotsys}) showing similar results, confirming the benefit of high-order SH features, in agreement with HEGNN~\cite{cen2024high} and prior analyses~\cite{joshi2023expressive}.

\subsection{RNA Property Prediction}
\label{sec:rna_property}

We evaluate DualEquiNet on RNA property prediction tasks because RNA molecules hold significant therapeutic potential, and accurate in-silico modeling can greatly reduce experimental costs and time~\cite{damase2021limitless, yazdani2024helm, prakash2024bridging}. RNA molecules pose unique modeling challenges due to their large size, complex secondary and tertiary structures involving both short-range and long-range interactions~\cite{sponer2018rna, xu2024beyond, xu2025harmony}, and the necessity for equivariant representations respecting rotations and translations~\cite{moskalev2024sehyena}. An effective RNA modeling approach must efficiently integrate both local atomic interactions and global structural dependencies.

We use three RNA datasets consisting of thousands of atoms: (i) \textbf{CovidVaccine}~\cite{wayment2022deep}, which provides nucleotide-level reactivity and degradation labels relevant for vaccine stability; (ii) \textbf{Ribonanza}~\cite{he2024ribonanza}, which includes nucleotide-level reactivity under different compounds to assess RNA folding and flexibility; and (iii) \textbf{Tc-ribo}~\cite{groher2018tuning}, which contains RNA-level regulatory behavior in response to tetracycline. 3D RNA structures are generated using RhoFold~\cite{shen2022e2efold} following prior works~\cite{moskalev2024sehyena}.
 
We follow the experimental protocol in~\cite{moskalev2024sehyena} to use eta-theta 
pseudotorsional backbone~\cite{wadley2007evaluating} as input and use Root Mean Squared Error (RMSE) as our performance metric. All models are evaluated using consistent train-validation-test splits (8:1:1), with hyperparameters selected based on the best validation performance, optimized via Optuna~\cite{akiba2019optuna} to ensure fair comparisons. 

As shown in Table~\ref{tab:rna_property}, DualEquiNet outperforms all baselines, reducing RMSE by 25.7\%, 33.5\%, and 8.4\% on CovidVaccine, Ribonanza, and Tc-ribo, respectively compared to best baseline. This highlights DualEquiNet’s strength in modeling complex RNA interactions, especially long-range dependencies, making it a strong candidate for RNA property prediction tasks.

\begin{table}[t]
\caption{Performance of RNA property prediction (RMSE$\pm$standard deviation (std.)). The best-performing model is highlighted in bold. DualEquiNet outperforms other baselines on each metric.}
\label{tab:rna_property}
\resizebox{\textwidth}{!}{%
\begin{tabular}{ccccccccc}
\toprule
\multirow{2}{*}{Method} & \multicolumn{4}{c}{\textbf{COVID}}                                             & \multicolumn{3}{c}{\textbf{Ribonanza}}                       & \textbf{Tc-Ribo}         \\
\cmidrule(lr){2-5} \cmidrule(lr){6-8} \cmidrule(lr){9-9}
                    & Reactivity      & pH10            & Mg pH10         & Avg.            & DMS             & 2A3             & Avg.            &     Factor      \\
\midrule
\multicolumn{1}{c|}{SchNet}          & 0.451\footnotesize±0.006 & 0.633\footnotesize±0.012 & 0.518\footnotesize±0.010 & 0.534\footnotesize±0.011 & 0.909\footnotesize±0.018 & 0.966\footnotesize±0.004 & 0.937\footnotesize±0.012 & 0.737\footnotesize±0.002 \\
\multicolumn{1}{c|}{EGNN}            & 0.453\footnotesize±0.006 & 0.634\footnotesize±0.013 & 0.520\footnotesize±0.010 & 0.535\footnotesize±0.011 & 0.912\footnotesize±0.017 & 0.968\footnotesize±0.005 & 0.940\footnotesize±0.012 & 0.720\footnotesize±0.022 \\
\multicolumn{1}{c|}{GVPGNN}            & 0.448{\footnotesize ±0.006} & 0.631{\footnotesize ±0.012} & 0.515{\footnotesize ±0.009} & 0.531{\footnotesize ±0.011} & 0.901{\footnotesize ±0.015} & 0.963{\footnotesize ±0.004} & 0.932{\footnotesize ±0.010} & 0.736{\footnotesize ±0.003} \\
\multicolumn{1}{c|}{TFN}               & 0.453{\footnotesize ±0.006} & 0.634{\footnotesize ±0.012} & 0.520{\footnotesize ±0.010} & 0.535{\footnotesize ±0.011} & 0.911{\footnotesize ±0.018} & 0.968{\footnotesize ±0.004} & 0.939{\footnotesize ±0.012} & 0.694{\footnotesize ±0.008} \\
\multicolumn{1}{c|}{FastEGNN}          & 0.390{\footnotesize ±0.019} & 0.582{\footnotesize ±0.029} & 0.463{\footnotesize ±0.027} & 0.478{\footnotesize ±0.031} & 0.808{\footnotesize ±0.022} & 0.869{\footnotesize ±0.015} & 0.839{\footnotesize ±0.020} & 0.731{\footnotesize ±0.006} \\
\multicolumn{1}{c|}{HEGNN}             & 0.450{\footnotesize ±0.005} & 0.632{\footnotesize ±0.012} & 0.517{\footnotesize ±0.009} & 0.533{\footnotesize ±0.011} & 0.904{\footnotesize ±0.015} & 0.964{\footnotesize ±0.004} & 0.934{\footnotesize ±0.010} & 0.729{\footnotesize ±0.007} \\
\multicolumn{1}{c|}{GotenNet}          & 0.388{\footnotesize ±0.002} & 0.582{\footnotesize ±0.011} & 0.455{\footnotesize ±0.008} & 0.475{\footnotesize ±0.009} & 0.832{\footnotesize ±0.017} & 0.882{\footnotesize ±0.009} & 0.857{\footnotesize ±0.013} & 0.718{\footnotesize ±0.002} \\
\multicolumn{1}{c|}{\textbf{DualEquiNet}}          & \textbf{0.272{\footnotesize ±0.004}} & \textbf{0.448{\footnotesize ±0.010}} & \textbf{0.340{\footnotesize ±0.005}} & \textbf{0.353{\footnotesize ±0.006}} & \textbf{0.611{\footnotesize ±0.025}} & \textbf{0.505{\footnotesize ±0.024}} & \textbf{0.558{\footnotesize ±0.012}} & \textbf{0.636{\footnotesize ±0.049}} \\ 
\bottomrule
\end{tabular}
}
\end{table}

\subsection{Protein and RNA Solvent-Accessible Surface Area (SASA) prediction}

\begin{wraptable}{r}{0.6\textwidth}
\vspace{-1em}
\caption{Performance on SASA Prediction (RMSE±std).}
\label{tab:sasa}
\begin{tabular}{ccc}
\toprule
    & RNASolo & mRFP Protein (×10)        \\
\midrule
SchNet     & 48.560\footnotesize±2.187      &     8.370\footnotesize±0.963      \\
EGNN       & 33.011\footnotesize±2.323      &     9.054\footnotesize±1.303      \\
GVPGNN     & 37.704\footnotesize±1.050      &     20.351\footnotesize±0.964      \\
TFN        & 51.003\footnotesize±1.214      &     20.176\footnotesize±1.277     \\
FastEGNN   & 37.251\footnotesize±1.028      &     12.282\footnotesize±1.310    \\
HEGNN      & 37.782\footnotesize±1.795      &     12.382\footnotesize±3.525     \\
GotenNet   & 34.966\footnotesize±1.671      &     8.743\footnotesize±1.558      \\
\textbf{DualEquiNet}   & \textbf{31.972\footnotesize±0.859}  & \textbf{5.959\footnotesize±1.160}   \\
\bottomrule
\end{tabular}
\end{wraptable}

Solvent-accessible surface area (SASA) measures how exposed atoms or residues are to solvent, reflecting a large biomolecule's 3D conformation and influencing interactions and stability~\cite{shrake1973environment, mishra2018role}. 
Accurate SASA prediction is crucial for understanding and engineering molecular behavior~\cite{stanton2022accelerating}.

For protein SASA prediction, we use the existing mRFP protein dataset from~\cite{stanton2022accelerating}, with 3D protein structures generated using AlphaFold\cite{jumper2021highly}. This dataset contains ground-truth SASA values, offering a reliable benchmark for model evaluation. Additionally, we introduce a new SASA benchmark dataset for RNA, sourced from RNASolo~\cite{adamczyk2022rnasolo}, which includes experimentally determined 3D structures of RNA molecules. During preprocessing, we choose RNA molecules with sequence lengths ranging from 10 to 400 and exclude RNA sequences with multiple chains or non-AUGC nucleotides. Ground-truth residue-level SASA values are computed using DSSR~\cite{lu2015dssr}. Together with the existing protein SASA benchmark, this new RNA dataset provides a comprehensive benchmark for assessing models’ ability to capture detailed structural representations of both RNA and protein.

The baselines and data splits are consistent with those in Section~\ref{sec:rna_property} and we use Optuna for optimal hyperparameter search. RMSE to the ground truth SASA values is reported.
Table~\ref{tab:sasa} presents performance results on the SASA benchmarks. DualEquiNet consistently achieves the lowest prediction error across both RNA and protein datasets, demonstrating its superior ability to capture and refine both local and global structural features through its dual-space hierarchical representations.

\subsection{RNA Torsion Angle Prediction}
\label{sec:exp_torsionangle}
Torsion angles define rotations around specific bonds, shaping the 3D conformation of biomolecules and influencing their secondary and tertiary structures, interactions, and functions. Accurate prediction of these angles is essential for modeling structural geometry. While sequence modeling techniques have been proposed for RNA torsion angle prediction~\cite{abir2024deeprna, bernard2025rna}, they rely solely on nucleotide sequences and ignore the 3D structural context. In contrast, torsion angles are fundamentally determined by spatial geometry (more details in Appendix~\ref{appen:datasets_detials_torsion}.

Similar to SASA experiments, the \textbf{TorsionAngle} dataset uses ground-truth RNA sequences and structures from RNASolo and torsion angles are computed using DSSR. Consistent with~\cite{abir2024deeprna}, we adopt the circular Mean Absolute Error (MAE) loss. To predict seven torsion angles ($\alpha, \beta, \gamma, \delta, \epsilon, \zeta, \chi$), the model outputs 14 values—$\sin$ and $\cos$ for each angle. The angle $\theta$ is recovered via $\theta = \arctan(\sin(\theta), \cos(\theta))$, and the circular MAE is computed as 
\begin{equation}
\label{eq:circular_mae}
\Delta \theta_i=\left|\theta_{\text{pred}, i}-\theta_{\text{true}, i}\right|, 
\quad
\operatorname{MAE}(\theta)=\frac{1}{N_{nt}} \sum_{i=1}^{N_{nt}} \min \left(\Delta \theta_i, 360^{\circ}-\Delta \theta_i\right),
\end{equation}
where $N_{nt}$ is the number of predicted nucleotides.

Results presented in Table \ref{tab:torsion} illustrate the performance of various models across different torsion angles prediction. Our DualEquiNet achieves the lowest average error and performs best for 5 out of 7 angles prediction. 
These findings highlight that DualEquiNet effectively learns 3D structural representations, enabling accurate prediction of torsion angles based on molecular geometry.

\begin{table}[t]
\caption{Performance on TorsionAngle dataset with 7 predicted angles. (Circular MAE ± std.). 
}
\label{tab:torsion}
\resizebox{\textwidth}{!}{%
\begin{tabular}{ccccccccc}
\toprule
         & \textbf{Avg.}                    & \textbf{$\alpha$}                     & \textbf{$\beta$}                     & \textbf{$\gamma$}                     & \textbf{$\delta$}                     & \textbf{$\epsilon$}                     & \textbf{$\zeta$}                     & \textbf{$\chi$}     \\
\midrule
SchNet   & 18.54\footnotesize±0.86  & 31.38\footnotesize±1.44  & 17.54\footnotesize±0.20  & 24.44\footnotesize±1.35  & 7.68\footnotesize±0.25  & 15.48\footnotesize±0.56  & 20.50\footnotesize±1.30  & 12.78\footnotesize±0.63    \\
EGNN     & 17.79\footnotesize±0.61 & 29.62\footnotesize±0.98      & 17.25\footnotesize±0.26     & 22.96\footnotesize±0.79      & 7.54\footnotesize±0.16       & 15.17\footnotesize±0.67        & 19.72\footnotesize±1.11     & 12.29\footnotesize±0.58    \\
GVPGNN   & 17.56\footnotesize±0.21 & 29.69\footnotesize±0.41      & 16.94\footnotesize±0.25     & 22.77\footnotesize±0.49      & 7.02\footnotesize±0.26       & 15.02\footnotesize±0.57        & 19.63\footnotesize±0.54     & 11.85\footnotesize±0.15    \\
TFN      & 21.97\footnotesize±0.45 & 36.52\footnotesize±0.62      & 18.19\footnotesize±0.40     & 27.63\footnotesize±0.66      & 11.63\footnotesize±0.43      & 17.64\footnotesize±0.79        & 26.59\footnotesize±0.77     & 15.57\footnotesize±0.74    \\
FastEGNN & 18.89\footnotesize±0.22 & 31.96\footnotesize±0.58      & 17.43\footnotesize±0.40     & 24.64\footnotesize±0.60      & 8.47\footnotesize±0.45       & 15.66\footnotesize±0.45        & 21.31\footnotesize±0.33     & 12.78\footnotesize±0.13    \\
HEGNN    & 18.41\footnotesize±0.22 & 31.12\footnotesize±0.59      & 17.49\footnotesize±0.21     & 23.78\footnotesize±0.69      & 7.91\footnotesize±0.66       & 15.35\footnotesize±0.46        & 20.54\footnotesize±0.18     & 12.67\footnotesize±0.26    \\
GotenNet & 16.27\footnotesize±1.19 & \textbf{26.88\footnotesize±1.57}      & 15.90\footnotesize±0.60     & \textbf{21.06\footnotesize±1.34}      & 6.72\footnotesize±0.35       & 13.85\footnotesize±0.64        & 17.74\footnotesize±1.63     & 10.96\footnotesize±0.84    \\
DualEquiNet & \textbf{15.87\footnotesize±2.03} & 26.89\footnotesize±3.00      & \textbf{15.48\footnotesize±1.17}     & 21.12\footnotesize±2.68      & \textbf{6.30\footnotesize±0.45}       & \textbf{13.68\footnotesize±1.29}        & \textbf{17.03\footnotesize±1.84}     & \textbf{10.60\footnotesize±1.32}   \\
\bottomrule
\end{tabular}
}
\end{table}

\subsection{Ablation and Neighborhood Dynamics: Importance of Dual-Space Learning}

To quantify the contributions of key components in DualEquiNet, we perform an ablation study across large-scale RNA property datasets. We evaluate four configurations: (i) the full \textbf{DualEquiNet} model; (ii) \textbf{w/o CSIP}, which replaces contextual structure-informed pooling (CSIP) with standard mean pooling; (iii) \textbf{w/o CSIP/Cross}, which disables cross-space message passing but retains dual-space representations; and (iv) \textbf{w/o CSIP/Cross/Dual}, a single-space Euclidean model. As shown in  Table~\ref{tab:ablation}, each component enhances performance, with CSIP particularly benefiting the Ribonanza and Tc-ribo datasets, and the dual-space mechanism notably improving performance on COVID. These results confirm that combining local and global geometric contexts through dual-space learning and cross-space communication is essential for generalization.

\begin{table}
\centering
\caption{Ablation Study on RNA Property datasets (RMSE±std.).}
\label{tab:ablation}
\begin{tabular}{lccc}
\toprule
Model & COVID $\downarrow$ & Ribonanza $\downarrow$ & Tc-Ribo $\downarrow$ \\
\midrule
\textbf{DualEquiNet} & \textbf{0.353±0.006} & \textbf{0.558±0.123} & \textbf{0.636±0.049} \\
w/o CSIP                 & 0.356\footnotesize±0.002 & 0.586\footnotesize±0.016 & 0.682\footnotesize±0.026 \\
w/o CSIP/Cross          & 0.356\footnotesize±0.013 & 0.591\footnotesize±0.081 & 0.698\footnotesize±0.010 \\
w/o CSIP/Cross/Dual    & 0.362\footnotesize±0.010 & 0.592\footnotesize±0.028 & 0.699\footnotesize±0.010 \\
\bottomrule
\end{tabular}
\end{table}

\subsection{Neighborhood Analysis}
\label{appen:neighbor_analysis}

To better understand how DualEqui captures both local and global geometric contexts, we analyze how neighborhoods evolve across layers in the EU and SH spaces. We use the experiment of DualEqui with 4 atom layers on the Covid Vaccine dataset as a case study. At each layer, EU and SH neighborhoods are dynamically redefined based on updated 3D coordinates and SH features, respectively. We compute 4 statistics: 

(1) the average distance of coordinates between each node and its EU neighbors; 
\begin{equation}
    dis_{EU} = \frac{1}{N_{atom}} \sum_{i=1}^{N_{atom}} \frac{1}{\left| \mathcal{N}_{EU}(i) \right|} \sum_{j \in \mathcal{N}_{EU}(i)} \lVert \boldsymbol{x}_i  - \boldsymbol{x}_j \rVert 
\end{equation}
(2) the average coordinates distance between each node and its SH neighbors; 
\begin{equation}
    dis_{SH} = \frac{1}{N_{atom}} \sum_{i=1}^{N_{atom}} \frac{1}{\left| \mathcal{N}_{SH}(i) \right|} \sum_{j \in \mathcal{N}_{SH}(i)} \lVert \boldsymbol{x}_i  - \boldsymbol{x}_j \rVert 
\end{equation}
(3) the average cosine similarity of SH features between each node and its EU neighbors; 
\begin{equation}
    cos_{EU} = \frac{1}{N_{atom}} \sum_{i=1}^{N_{atom}} \frac{1}{\left| \mathcal{N}_{EU}(i) \right|} \sum_{j \in \mathcal{N}_{EU}(i)} \frac{ \boldsymbol{r}_{i} \cdot \boldsymbol{r}_{j}}{\|\boldsymbol{r}_{i}\| \cdot \|\boldsymbol{r}_{j}\| }
\end{equation}
(4) the average cosine similarity of SH features between each node and its SH neighbors. 
\begin{equation}
    cos_{SH} = \frac{1}{N_{atom}} \sum_{i=1}^{N_{atom}} \frac{1}{\left| \mathcal{N}_{SH}(i) \right|} \sum_{j \in \mathcal{N}_{SH}(i)} \frac{ \boldsymbol{r}_{i} \cdot \boldsymbol{r}_{j}}{\|\boldsymbol{r}_{i}\| \cdot \|\boldsymbol{r}_{j}\| }
\end{equation}

\begin{figure}[h]
    \centering
    \begin{subfigure}{0.4\textwidth}
    \label{fig:layer_dist}
        \includegraphics[width=\linewidth]{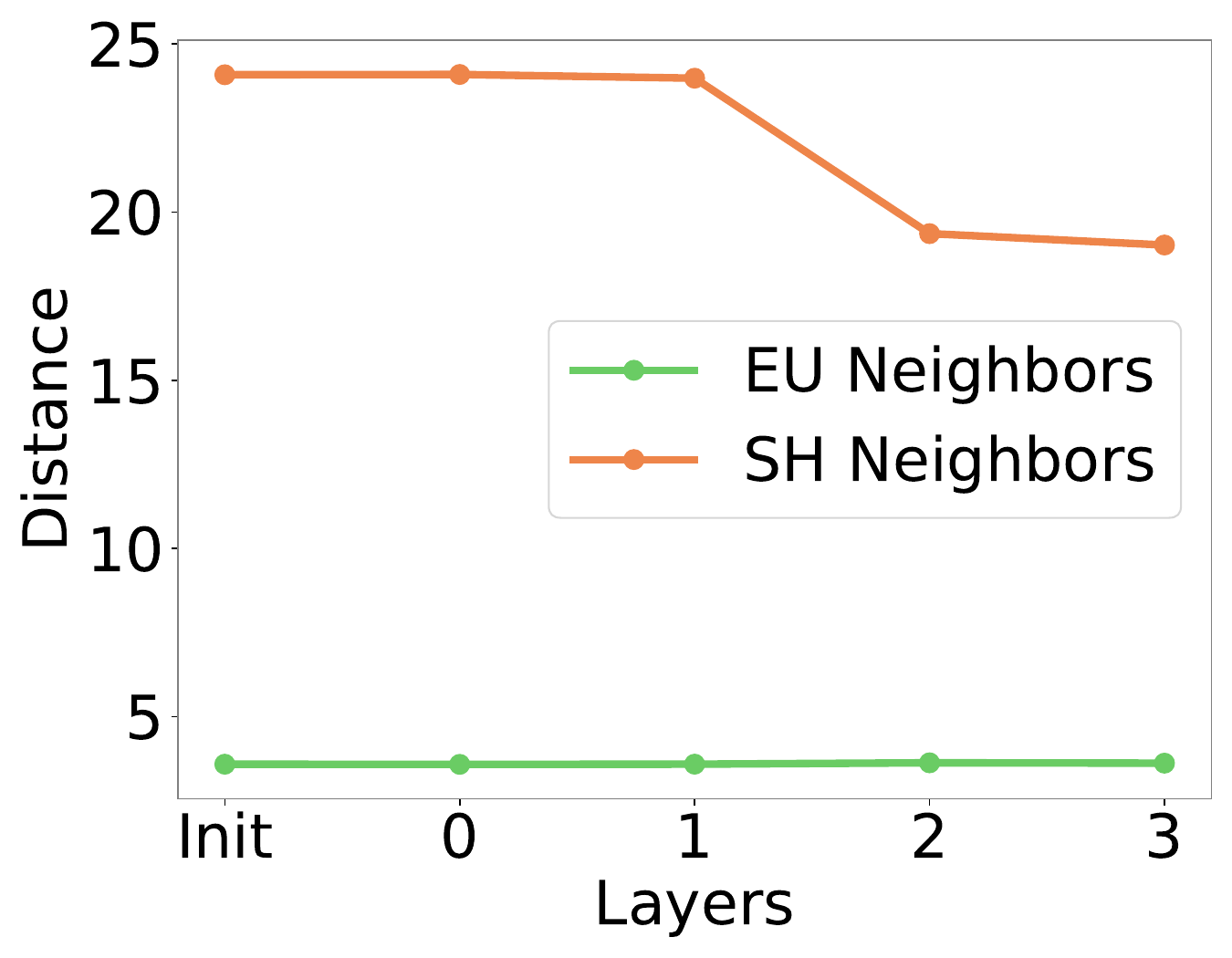}
    \end{subfigure}
    \hfill
    \begin{subfigure}{0.4\textwidth}
    \label{fig:layer_similarity}
        \includegraphics[width=\linewidth]{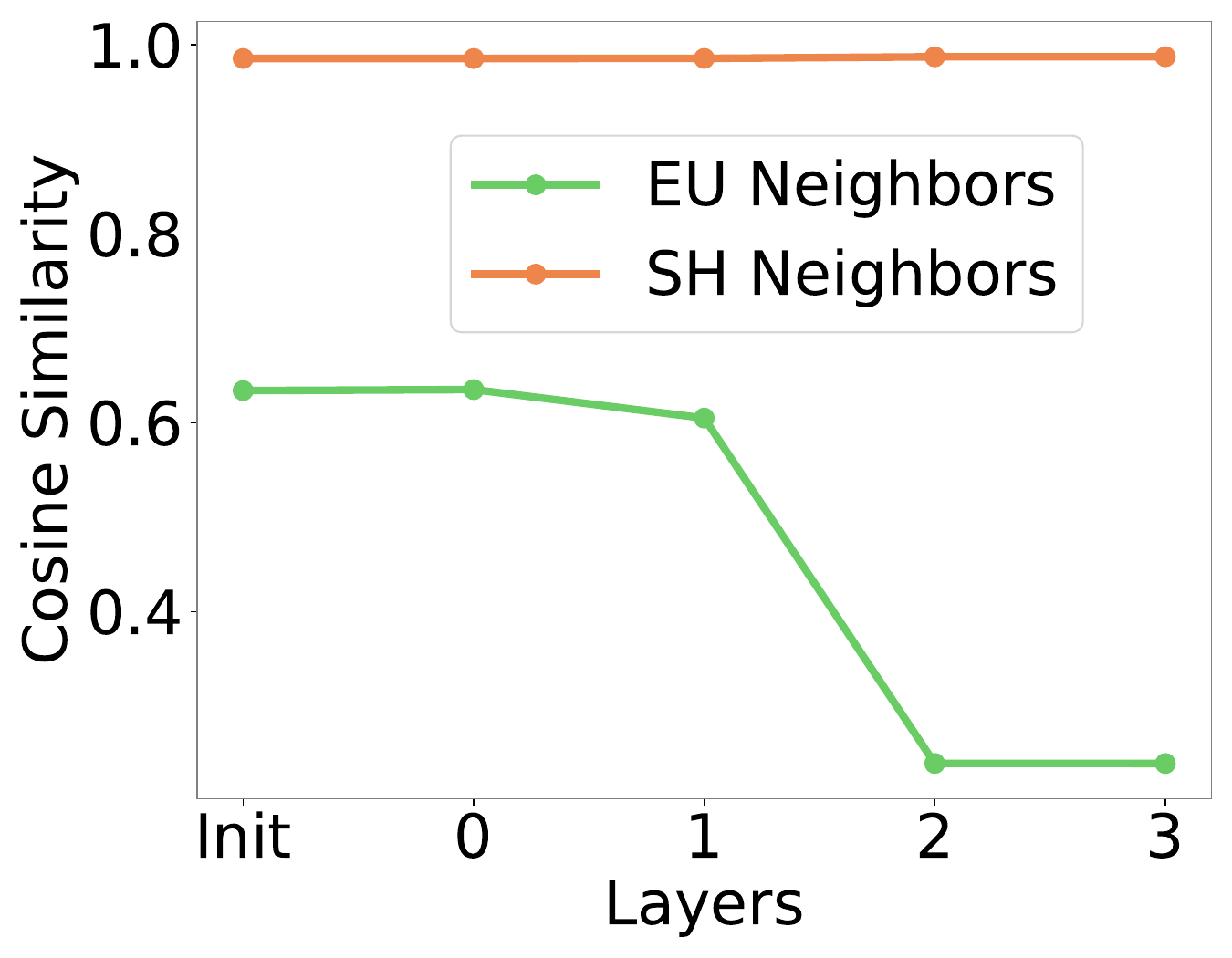}
    \end{subfigure}
    \caption{(a) Left: Average distance variation of EU and SH neighborhoods across layers. (b) Right: Average similarity variation of EU and SH neighborhoods across layers.}
    \label{fig:dist_similarity}
\end{figure}

The distances of coordinates are plotted in Fig.~\ref{fig:dist_similarity}(a), and the cosine similarity of SH features are plotted in Fig.~\ref{fig:dist_similarity}(b). 
Results reveal that the average distance of EU neighbors ($dis_{EU}$) and the SH features cosine similarity of SH neighbors ($cos_{SH}$) remain stable across layers. This behavior is expected since EU neighborhoods are constructed using a fixed cutoff $d_{EU}$ based on spatial proximity, while SH neighborhoods are selected using a threshold $d_{SH}$ on SH feature cosine similarity. In this experiment, we use $d_{EU} = 4.8$ and $d_{SH} = 0.97$.

In contrast, we observe clear trends in the other two metrics. The average coordinates distance of SH neighbors ($dis_{SH}$) decreases significantly over layers (Fig.\ref{fig:dist_similarity}(a), orange line), indicating that nodes with similar SH features—which initially may be far apart in EU space—gradually become closer in Euclidean distance. This trend reflects the impact of the cross-space update in Eq.\ref{eq:delta_x_e} and~\ref{eq:delta_r_sh}. When we update the SH features $\boldsymbol{r}$, we also consider the EU coordinates, which brings distant SH neighbors closer in EU space. In practice, we can imagine the scenario in biomolecules where spatially distant residues can interact through folding.

Meanwhile, the SH features cosine similarity of EU neighbors ($cos_{EU}$) drops across layers (Fig.\ref{fig:dist_similarity}(b), green line), implying that while EU neighbors may remain close in EU space, their SH features increasingly diverge. According to Eq.\ref{eq:init_sh_neigh}, the SH neighbors are initialized according to the EU neighbors at first and therefore have a higher similarity score. But after several layers, the SH features are updated to differentiate local nodes based on their broader geometric context or functional roles. This is relevant in structured biomolecules where nearby atoms may belong to different residues or to distinct structural regions.

Thus, the observed dynamics validate the model's dual-space architecture: SH space helps bridge distant but functionally related parts of the molecule, while EU space preserves local geometry but allows for semantic diversity.

More experiments about the time complexity analysis and expressivity on symmetric structures are in Appendix~\ref{appen:exp_details}.

\section{Conclusion}
\label{sec:conclusion}

We introduce DualEquiNet, a novel framework for modeling large biomolecules by hierarchically integrating Euclidean and spherical harmonics representations across atomic, residue (nucleotide/amino acid), and molecular levels. DualEquiNet is E(3) equivariant and effectively captures both local and global structural dependencies. Extensive experiments on real-world datasets demonstrate state-of-the-art performance over existing geometric GNNs, highlighting the benefits of dual-space representations and hierarchical learning for large biomolecular modeling. Although DualEquiNet is effective across diverse tasks, its performance may depend on the quality of input 3D structures, which can vary across datasets and affect generalization. Future work will explore integrating structure prediction into the pipeline and extending the framework to broader molecular applications.

\newpage
\bibliography{ref}
\bibliographystyle{plain}

\appendix
\newpage


\section{Related Work}
\label{appen:related}

\subsection{Invariant GNNs}
Graph Neural Networks (GNNs) learn from graph-structured data by aggregating information from local neighborhoods~\cite{wu2020comprehensive, dai2022comprehensive}. Invariant GNNs learn molecular representations using rotation-invariant geometric features, ensuring predictions remain unchanged under rigid transformations. SchNet employs continuous-filter convolutions over interatomic distances to model molecular properties effectively~\cite{schutt2017schnet}. DimeNet extends SchNet by integrating angular features, explicitly considering bond angles to capture directional information~\cite{gasteiger2020directional, gasteiger2020fast, klicpera2020directional}. GemNet further enhances expressiveness by utilizing multi-hop distance–angle interactions, implicitly encoding higher-order geometric features~\cite{gasteiger2021gemnet}. SphereNet incorporates a comprehensive set of invariant scalars, including distances, angles, and torsion angles, to robustly represent molecular structures~\cite{liu2022spherical}. 

Despite their effectiveness, invariant GNNs inherently lose crucial directional and orientational information, motivating the development of equivariant models capable of explicitly encoding full geometric contexts.

\subsection{Scalarization Equivariant GNNs}
Scalarization equivariant graph neural networks maintain E(3) equivariance by explicitly handling vector-valued features through scalarized updates. EGNN~\cite{satorras2021n} leverages pairwise distances to compute scalar-valued messages, enabling direct equivariant updates to atomic positions without high-order tensor operations. GVPGNN~\cite{jing2020learning} introduces geometric vector perceptrons, managing scalar and vector embeddings separately and using scalar gating to achieve SO(3)-equivariant transformations. PaiNN~\cite{schutt2021equivariant} similarly utilizes scalarized vector updates, applying distance-based filters to coordinate features, thus ensuring consistent equivariant message passing. 

These models share the common technique of scalarization, employing invariant scalars to guide separate, equivariant updates of vector or positional features. However, scalarization methods inherently lack the expressive power provided by high-degree spherical tensor representations, potentially limiting their ability to capture complex rotational relationships and subtle structural differences in molecular systems~\cite{joshi2023expressive}.

\subsection{Higher-degree Equivariant GNNs}
\paragraph{CG-based Tensor Product}  Several equivariant GNNs achieve expressivity by using high-order irreducible representations combined via Clebsch–Gordan tensor products. Tensor Field Networks~\cite{thomas2018tensor} (TFN) introduced this paradigm by building SE(3)-equivariant convolutional layers from CG-based feature tensor products, enabling rotation-equivariant learning with theoretically maximal expressiveness at a significant computational cost. The SE(3)-Transformer~\cite{fuchs2020se} extended TFN by incorporating a self-attention mechanism on irreducible feature spaces, further improving representation power. SEGNN~\cite{brandstetter2021geometric} construct streeable MLPs by using the tensor product so that formalize an equivariant message aggregation.MACE~\cite{batatia2022mace} uses high-order tensor features to form a complete basis of scalar invariants through the Atomic Cluster Expansion, capturing arbitrary many-body interactions. However, it also incurs significant overhead due to numerous CG tensor product computations. 

\paragraph{Efficient Implementation}  Notably, the reliance on full CG tensor products leads to steep scaling in complexity (e.g. $O(L^6)$ for degree $L$ features). Due to the high computation cost of CG-based tensor product, more recent approaches aim to retain high-degree expressivity in a more efficient way. HEGNN~\cite{cenhigh} incorporate higher-degree spherical harmonics features but using EGNN-like inner-product scalarization trick to handle cross-degree interactions. Similarly, GotenNet~\cite{aykent2025gotennet} introduces high-order spherical tensor in a unified attention architecture that eschews explicit CG transforms; instead, it updates features via inner products on steerable tensors. Our DualEqui also follows this paradigm: it combines spherical harmonic feature channels with dual scalarization steps to capture rich 3D geometric information while keeping computations tractable.

\subsection{Long-Range Dependencies in GNNs}

Learning long-range dependencies is a key challenge in graph machine learning, arising from the fundamental trade-off between under-reaching (i.e., insufficient GNN layers leading to small receptive fields) and over-smoothing (i.e., excessive layers causing feature homogenization)~\cite{chen2020simple,nguyen2023revisiting}. A variety of benchmarks~\cite{dwivedi2022long,tonshoff2024where,zhou2025glora} have been proposed to study this problem, along with several representative solutions, including Implicit Graph Neural Networks (IGNNs) and Graph Transformers (GTs). IGNNs define message-passing through fixed-point equations~\cite{gu2020implicit,liu2022mgnni} or unfolded optimization steps~\cite{chen2022optimization,baker2023implicit}, allowing stable convergence to an equilibrium state even with infinite layers. Efforts to improve their scalability have focused on accelerating convergence~\cite{chen2022efficient,baker2023implicit} and efficient neighbor sampling~\cite{liu2024scalable,li2022unbiased}, making IGNNs more practical for large-scale applications. GTs leverage global attention mechanisms~\cite{wu2021representing,ying2021transformers,rampasek2022GPS} to capture long-range dependencies while incorporating structural encodings to preserve graph properties. Recent works have improved their scalability~\cite{wu2022nodeformer,wu2023sgformer,luo2024enhancing} and extended their applicability to directed graphs~\cite{geisler2023transformers,luo2023transformers}, broadening their impact in graph learning.


\section{Basics of Group Theory, Equivariance, and Spherical Harmonics}
\label{appen:math}

In this section, we provide an introduction to the mathematical background used in this paper. For readers interested in more details, we refer them to the Guide to Geometric GNNs~\cite{duval2023hitchhiker} and the e3nn documentation~\cite{geiger2022e3nn}.

\subsection{Groups and Symmetry}
A group is an algebraic structure consisting of a set $\mathfrak{G}$ together with a binary operation (composition) that satisfies four fundamental axioms:
\begin{itemize}
    \item Closure: For any $\mathfrak{a}, \mathfrak{b} \in \mathfrak{G}$, the composite $\mathfrak{a}\cdot \mathfrak{b}$ is also in $\mathfrak{G}$.
    \item Identity: There is an identity element $\mathfrak{e} \in \mathfrak{G}$ such that $\mathfrak{e} \cdot \mathfrak{a} = \mathfrak{a} \cdot \mathfrak{e} = \mathfrak{a}$ for every $\mathfrak{a} \in \mathfrak{G}$.
    \item Inverse: For each $\mathfrak{a} \in \mathfrak{G}$, there exists an inverse element $\mathfrak{a}^{-1} \in \mathfrak{G}$ with $\mathfrak{a}\cdot \mathfrak{a}^{-1} = \mathfrak{a}^{-1}\cdot \mathfrak{a} = \mathfrak{e}$.
    \item Associativity: For all $\mathfrak{a}, \mathfrak{b}, \mathfrak{c} \in \mathfrak{G}$, we have $(\mathfrak{a} \cdot  \mathfrak{b}) \cdot \mathfrak{c} = \mathfrak{a} \cdot (\mathfrak{b} \cdot \mathfrak{c})$.
\end{itemize}

Groups are often used to describe symmetries of objects or spaces. For example, the set of all rotations about a fixed point forms a group, as does the set of all translations in the plane. The symmetries of an object form a group under composition of transformations. 

\textbf{The Euclidean Group E(3):} In geometric deep learning, an important example is the Euclidean group in three dimensions, E(3), which captures the symmetries of 3D Euclidean space (i.e., all rigid motions in $\mathbb{R}^3$). An element of E(3) can be represented as a pair $(\mathbf{R}, \mathbf{t})$ where $R$ is a $3\times3$ rotation matrix and $\mathbf{t}\in\mathbb{R}^3$ is a translation vector. Composition in E(3) is done by composing rotations and adding translations: $(\mathbf{R}_1,\mathbf{t}_1)\cdot(\mathbf{R}_2,\mathbf{t}_2) = (\mathbf{R}_1\mathbf{R}_2,; \mathbf{R}_1\mathbf{t}_2 + \mathbf{t}_1)$. This group is essentially the semi-direct product of translations and rotations. Intuitively, applying an E(3) transformation to a point $\boldsymbol{x}\in\mathbb{R}^3$ yields $\mathbf{R}\boldsymbol{x} + \mathbf{t}$, which is just a rotated and translated version of the point. E(3) is the symmetry group of any rigid 3D object, meaning that if an object (or a dataset) is moved or rotated in space, it remains essentially the “same” object – a concept captured by invariance or equivariance (discussed in B.2). Many physical or geometric tasks exhibit E(3) symmetry. Hence, incorporating E(3) into machine learning models is a powerful inductive bias~\cite{bronstein2021geometric, kondor2018generalization}.

\textbf{The Rotation Group SO(3):} The special orthogonal group SO(3) is the subgroup of E(3) consisting of rotations about the origin (with no translations). Formally, SO(3) is the set of all $3\times3$ orthogonal matrices with determinant 1. Every element $\mathbf{R}\in\textrm{SO}(3)$ represents a rotation in 3D space. SO(3) plays a central role in describing angular symmetries. For example, the orientation of an object in space can be changed by an element of SO(3). SO(3) is non-abelian (the order of rotations matters), unlike the translation sub-group (which is abelian). In many applications (like molecular modeling or 3D vision), only rotations need to be considered (e.g., when comparing shapes regardless of orientation), making SO(3) a crucial symmetry group. Models that respect SO(3) symmetry avoid to learn the same feature in all rotated forms and group-theoretic approaches provide a unified framework to incorporate such symmetries into neural networks.

\subsection{Equivariance and Invariance in Geometric Learning}
Equivariance to Euclidean transformations is a desirable property in geometric learning tasks involving 3D molecular structures. Specifically, a function \( f: \mathcal{X} \rightarrow \mathcal{Y} \) is said to be \textit{equivariant} with respect to a group \( \mathfrak{G} \) (e.g., the Euclidean group \(\mathrm{E}(3)\)) if, for any transformation $\mathfrak{g} \in \mathfrak{G}$ and any input $\boldsymbol{x} \in \mathcal{X}$, there exists a corresponding transformation \( \mathfrak{g} \) acting on the output space such that
\begin{equation}
\label{eq:definition_equivariance}
    f(\mathfrak{g} \cdot \boldsymbol{x}) = \mathfrak{g} \cdot f(\boldsymbol{x}).
\end{equation}
For \(\mathrm{E}(3)\), this means that if the input (e.g., atomic coordinates or features) is rotated, translated, or reflected, the output of the model transforms in a consistent and predictable manner. This property ensures that the model's predictions are physically meaningful and invariant to arbitrary choices of coordinate systems—a key requirement for tasks involving 3D biomolecular data.

To achieve rotational equivariance in 3D, spherical harmonics are commonly used to represent directional features on the unit sphere. Spherical harmonics form an orthonormal basis for square-integrable functions defined on the sphere and serve as the irreducible representations of the rotation group \(\mathrm{SO}(3)\). A key property of these representations is their equivariant transformation under rotations. For a unit direction vector \(\hat{\boldsymbol{x}}\) and a rotation matrix \(\mathbf{R} \), the spherical harmonics basis of degree \(l\) transforms according to:
\begin{equation}
Y^{(l)}(\mathbf{R} \hat{\boldsymbol{x}}) = \mathcal{D}^{(l)}(\mathbf{R}) \, Y^{(l)}(\hat{\boldsymbol{x}}),
\end{equation}
where \(Y^{(l)}(\cdot)\) is the spherical harmonics of degree \(l\), and \(\mathcal{D}^{(l)}(\mathbf{R}) \in \mathbb{R}^{(2l+1) \times (2l+1)}\) is the Wigner \(D\)-matrix corresponding to rotation \(\mathbf{R}\). This property enables the construction of rotation-equivariant functions by combining SH features with these bases, making them particularly suitable for encoding angular dependencies in molecular systems. We introduce more background in Appendix~\ref{appen:exp_details}.

\textbf{Invariance vs. Equivariance:} Invariant functions are a special case of equivariant functions. We say $f: \mathcal{X}\to \mathcal{Y}$ is invariant to $\mathfrak{G}$ if $f(\mathfrak{g}\cdot \boldsymbol{x}) = f(\boldsymbol{x})$ for all $\mathfrak{g}\in \mathfrak{G}$, i.e. the output is unchanged by the transformation. This corresponds to equation~\ref{eq:definition_equivariance} where the $\mathfrak{G}$-action on $\mathcal{Y}$ is trivial (identity for all $\mathfrak{g}$). Invariance discards all information about the transformation (losing equivariant detail), whereas equivariance preserves information in a structured way. Generally, equivariance is more powerful than invariance when building deep representations because it allows higher layers to reason about what transformation occurred, not just that something is symmetric.

\subsection{SH Space Captures Global Information}
\label{sec:sh_explain}

Spherical Harmonics (SH) provide an orthonormal basis for functions defined on a sphere, similar to how the Fourier series represents periodic functions. In this section, we draw a parallel between Fourier decomposition in one dimension and SH expansion on the sphere, highlighting how increasing the expansion order refines function approximation and why SH captures global structural information.

\subsubsection{Fourier Series vs. Spherical Harmonics}
A well-known tool for analyzing periodic functions is the Fourier series, which decomposes a function into a sum of sinusoidal components. Given a periodic function $ f(x) $ defined on $ x \in [-\pi, \pi] $, its Fourier series expansion is:
\begin{equation}
    f(x) = a_0 + \sum_{n=1}^{\infty} \left( a_n \cos(n x) + b_n \sin(n x) \right),
\end{equation}
where the coefficients are computed as:
\begin{equation}
    a_n = \frac{1}{\pi} \int_{-\pi}^{\pi} f(x) \cos(n x) \, dx, \quad
    b_n = \frac{1}{\pi} \int_{-\pi}^{\pi} f(x) \sin(n x) \, dx.
\end{equation}
By truncating the expansion at order $ N $, we obtain an approximation:
\begin{equation}
    f_{\text{approx}}(x) = a_0 + \sum_{n=1}^{N} \left( a_n \cos(n x) + b_n \sin(n x) \right).
\end{equation}
As $ N \to \infty $, the approximation converges to the true function $ f(x) $. In this representation:
\begin{itemize}[leftmargin=25pt]
    \item \textbf{Low-frequency terms} ($ n = 1,2 $) capture smooth, large-scale structures.
    \item \textbf{High-frequency terms} ($ n \gg 2 $) refine the function by encoding local variations.
\end{itemize}

Analogously and Intuitively, SH expansion generalizes this idea to functions defined on the sphere. For a function $ f(\theta, \phi) $ defined on the unit sphere, the \textbf{SH decomposition} is:
\begin{equation}
    f(\theta, \phi) = \sum_{l=0}^{\infty} \sum_{m=-l}^{l} c_m^{(l)} Y_m^{(l)}(\theta, \phi),
\end{equation}
where the coefficients are obtained via:
\begin{equation}
    c_m^{(l)} = \int_{\mathbb{S}^2} f(\theta, \phi) Y_m^{(l)}(\theta, \phi) \sin\theta \, d\theta \, d\phi .
\end{equation}
Truncating the expansion at order $ L $ provides an approximation:
\begin{equation}
    f_{\text{approx}}(\theta, \phi) = \sum_{l=0}^{L} \sum_{m=-l}^{l} c_m^{(l)} Y_m^{(l)}(\theta, \phi).
\end{equation}
As $ L \to \infty $, this converges to the true function $ f(\theta, \phi) $. The role of $ L $ in SH is similar to $ N $ in Fourier analysis:
\begin{itemize}[leftmargin=25pt]
    \item \textbf{Low-degree terms} ($ l = 0,1,2 $) capture smooth, global variations.
    \item \textbf{High-degree terms} ($ l \gg 2 $) encode fine-grained local details.
\end{itemize}

\subsubsection{Why Spherical Harmonics Capture Global Information}
Spherical harmonics enable long-range information encoding because they provide a global basis rather than relying on local neighborhoods. Unlike message-passing GNNs, which aggregates information only from spatially close neighbors, SH representations allow direct interactions between distant nodes that share similar SH coefficients.

\begin{figure}[H]
    \centering
    \includegraphics[width=\linewidth]{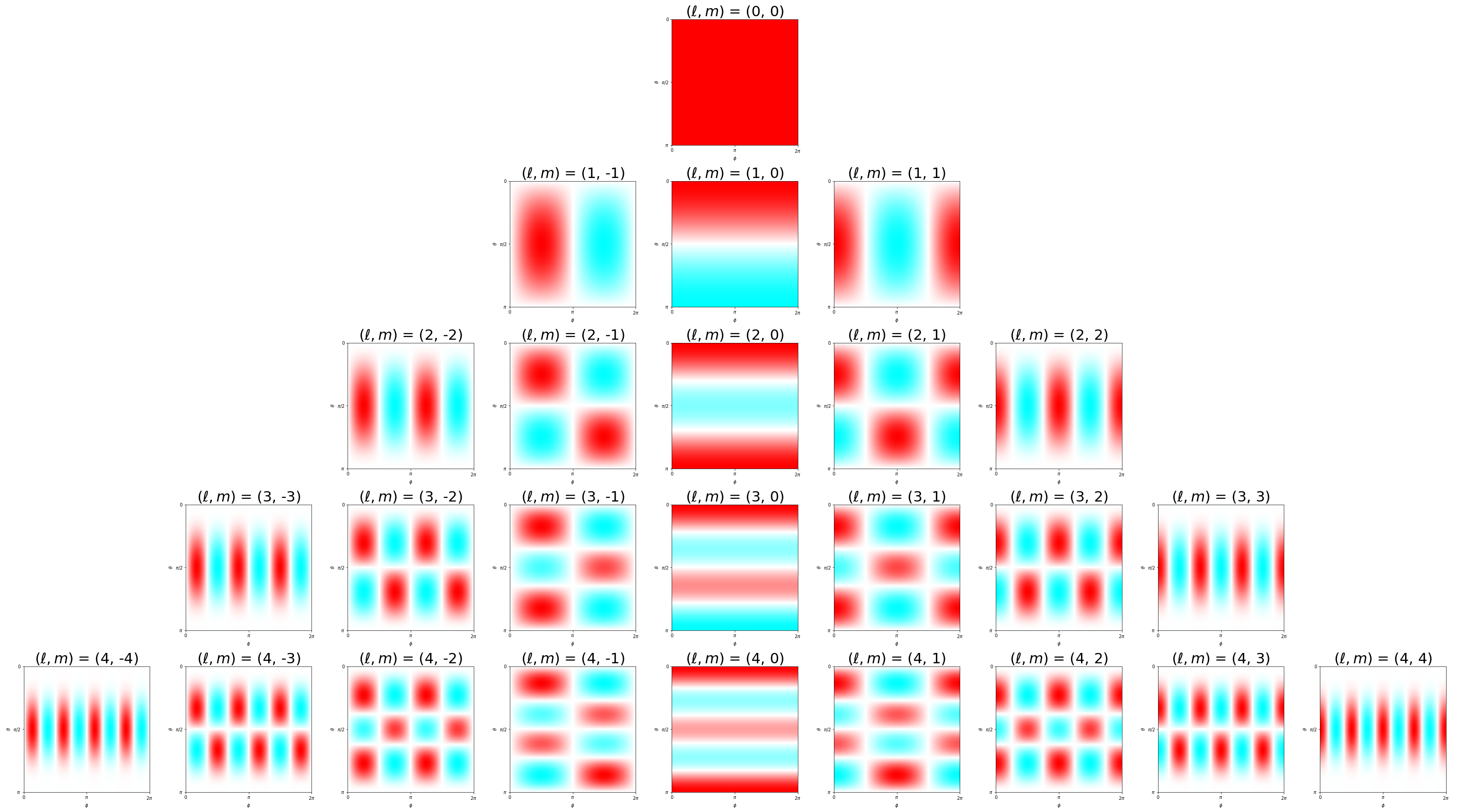}
    \caption[Visualization of real Spherical Harmonics table]{Visualization of real Spherical Harmonics table.\footnotemark}
    \label{fig:sh_table}
\end{figure}
\footnotetext{Image source: \href{https://en.wikipedia.org/wiki/Table_of_spherical_harmonics\%23Visualization_of_real_spherical_harmonics}{Wikipedia}}

Figure~\ref{fig:sh_table} provides a visualization of the first few real-valued Spherical Harmonics. The figure illustrates how the functions oscillate across different degrees. Lower-degree harmonics ($l$) exhibit slower variations, while higher-degree harmonics demonstrate more rapid oscillations along the axes, reflecting their ability to capture different scales of geometric features. To be more specific, when
\begin{itemize}[leftmargin=25pt]
    \item \textbf{$ l = 0 $ (Monopole)}: Captures the global molecular shape.
    \item \textbf{$ l = 1 $ (Dipole)}: Represents molecular orientation.
    \item \textbf{$ l = 2 $ (Quadrupole)}: Encodes bond angles and torsional interactions.
    \item \textbf{Higher $ l $ values}: Capture localized atomic-scale interactions.
\end{itemize}

Since low-degree SH functions change gradually, functions dominated by low-degree SH terms exhibit long-range correlations, as information propagates smoothly across extended regions of the domain.

\paragraph{Mathematical Justification.}
The SH functions are eigenfunctions of the Laplace-Beltrami operator:

\begin{equation}
    \Delta_{\mathbb{S}^2} Y_m^{(l)} = -l(l+1) Y_m^{(l)}.
\end{equation}

Since the eigenvalues $ -l(l+1) $ control how rapidly the functions oscillate, low-degree SH components ($ l = 0,1,2 $) vary slowly and thus retain long-range structural features.

\paragraph{SH-Based Neighborhoods Enable Long-Range Dependencies.}
In our framework, nodes interact based on the similarity of their spherical harmonics (SH) feature representations, independent of their Euclidean spatial distance. This enables two spatially distant nodes to have strong interactions if they share similar local geometric environments (e.g., comparable bond angles and torsions with their respective neighbors), as captured by their SH representations.

Traditional message passing in molecular graphs is inherently local, where information propagates only between spatially adjacent nodes. To capture long-range dependencies, these methods require multiple message passing steps, which can be computationally expensive and may lead to information dilution.

Our SH-based approach naturally encodes global structural dependencies while maintaining computational efficiency. This is possible because low-degree spherical harmonics have broad spatial support, creating smooth, global basis functions that span large regions of the molecular structure. As a result, nodes with similar SH coefficients maintain strong correlations regardless of their spatial separation.

This framework effectively captures long-range molecular interactions without requiring explicit long-range message passing, as the similarity in SH representations automatically establishes meaningful connections between distant nodes that share similar local geometric environments.

\section{Proof of Equivariance}
\label{appen:proofs}

In this section, we formally demonstrate the E(3)-equivariance property of our proposed DualEqui model. Specifically, we begin by proving the invariance property of the message functions, showing that they are built upon fundamental invariant quantities. Next, we analyze the forward computation of the DualEqui model and establish the equivariance property.

\paragraph{Notations.} In this proof, we let $\boldsymbol{X} \in \mathbb{R}^{N\times 3}$ represent the set of all input 3D atomic coordinates, where each row $\boldsymbol{x}_i\in\mathbb{R}^3$ (for $i\in\{1,\cdots,N\}$) corresponds to one specific atomic position within $\boldsymbol{X}$. We represent an arbitrary E(3) transformation using a rotation matrix $\mathbf{R} \in \mathbb{R}^{3\times3}$ and a translation vector $\mathbf{t}\in\mathbb{R}^3$. 

For convenience, we let $z$ be an arbitrary intermediate variable in our model computation (e.g., message $m_{EU,ij}$ in Eq.~\eqref{eq:eu_space_msg}). Occasionally, we adopt a functional notation, expressing it as a function of all input 3D coordinates, i.e., $z(\boldsymbol{X})$. This allows us to denote the model output under E(3) transformations as $z(\mathbf{R} \boldsymbol{X}+\mathbf{t})$ and verify whether equivariance or invariance holds. By default, we consider the original non-functional form of intermediate variables $z$ as shorthand notation for $z(\mathbf{R}\boldsymbol{X}+\mathbf{t})$, where $\mathbf{R} = \mathbf{I}$ and $\mathbf{t}=0$.

\subsection{Basic Invariant Quantities}

The equivariance of the DualEqui model is grounded in the invariance of several fundamental quantities. Our message functions in Eq.~\eqref{eq:eu_space_msg} and Eq.~\eqref{eq:sh_space_msg} are computed using two basic invariant quantities, whose invariance is shown below.

\paragraph{Euclidean Distance $\|\boldsymbol{x}_{ij}\|$.} Let $\boldsymbol{x}_i,\boldsymbol{x}_j$ be any pair of 3D coordinate vectors. When an E(3) transformation is applied to both $\boldsymbol{x}_i$ and $\boldsymbol{x}_j$, the Euclidean distance is invariant due to the orthogonality of rotation matrices:
\begin{align*}
    \| (\mathbf{R} \boldsymbol{x}_i + \mathbf{t}) - (\mathbf{R} \boldsymbol{x}_j + \mathbf{t}) \| = & ~ \| \mathbf{R} (\boldsymbol{x_i} - \boldsymbol{x}_j) \| \\
        = & ~ \sqrt{(\boldsymbol{x_i} - \boldsymbol{x}_j)^\top \mathbf{R}^\top \cdot \mathbf{R} (\boldsymbol{x}_i - \boldsymbol{x}_j)} \\
        = & ~\| \boldsymbol{x}_i - \boldsymbol{x}_j\| \\
        = & ~ \| \boldsymbol{x}_{ij}\|.
\end{align*}

\paragraph{Spherical Harmonics Feature $\boldsymbol{r}_i \odot \boldsymbol{r}_j$.} To show the invariance of $\boldsymbol{r}_i \odot \boldsymbol{r}_j$, we first consider the equivariance of an arbitrary $\boldsymbol{r}_i^l$ for node $i\in\{1,\cdots,N\}$ and order $l\in\{0,\cdots,l_{max}\}$ under E(3) transformations. Specifically, we have:
\begin{align}\label{eq:eqvar_r_il}
    \boldsymbol{r}_i^{(l)}(\mathbf{R}\boldsymbol{X}+\mathbf{t}) = & ~ \frac{1}{|\mathcal{N}_{EU}(i)|} \sum_{j \in \mathcal{N}_{EU}(i)} \phi_0 \left( \left[h_i, h_j, \left\|\mathbf{R}\boldsymbol{x}_{ij} \right\| \right] \right) \  \hat{Y}^l \left( \mathbf{R}\boldsymbol{\hat{x}}_{ij} \right) \notag \\
    = & ~ \frac{1}{|\mathcal{N}_{EU}(i)|} \sum_{j \in \mathcal{N}_{EU}(i)} \phi_0 \left( \left[h_i, h_j, \left\|\boldsymbol{x}_{ij} \right\| \right] \right) \  D^{(l)}(\mathbf{R})\hat{Y}^l \left( \boldsymbol{\hat{x}}_{ij}\right) \notag \\
    = & ~ D^{(l)}(\mathbf{R}) \cdot \left(\frac{1}{|\mathcal{N}_{EU}(i)|}\sum_{j \in \mathcal{N}_{EU}(i)} \phi_0 \left( \left[h_i, h_j, \left\|\boldsymbol{x}_{ij} \right\| \right] \right) \ \hat{Y}^l \left( \boldsymbol{\hat{x}}_{ij}\right)\right) \notag \\ 
    = & ~ D^{(l)}(\mathbf{R}) \boldsymbol{r}_i^{(l)}, 
\end{align}
where $D^{(l)}(\mathbf{R})$ denotes the $l$-th order Wigner-D matrix, which denotes the irreducible representation of the rotation group SO(3). Specifically, all the Wigner-D matrices are unitary, i.e., $D^{(l)}(\mathbf{R})^\top D^{(l)}(\mathbf{R}) = \mathbf{I}$. 

Thus, since Eq.~\eqref{eq:sh_feat_all_order} already establishes that
\begin{align*}
    \boldsymbol{r}_i = [\boldsymbol{r}^{(0)}_i, \boldsymbol{r}^{(1)}_i, \cdots, \boldsymbol{r}^{(l_{max})}_i].
\end{align*}

We can then directly obtain
\begin{align*}
    \boldsymbol{r}_i(\mathbf{R}\boldsymbol{X}+\mathbf{t}) = & ~ [D^{(0)}(\mathbf{R})\boldsymbol{r}^{(0)}_i, D^{(1)}(\mathbf{R})\boldsymbol{r}^{(1)}_i, \cdots, D^{(l_{max})}(\mathbf{R})\boldsymbol{r}^{(l_{max})}_i].
\end{align*}

Therefore, Since the Wigner-D matrices are unitary, the invariance of $\boldsymbol{r}_i \odot \boldsymbol{r}_j$ follows:
\begin{align*}
    & ~ \boldsymbol{r}_i(\mathbf{R}\boldsymbol{X} +\mathbf{t}) \odot \boldsymbol{r}_j(\mathbf{R}\boldsymbol{X} +\mathbf{t}) \\
    = & ~ \left[\left(\boldsymbol{r}^{(0)}_iD^{(0)}(\mathbf{R})\right)^\top 
 \cdot \left(D^{(0)}(\mathbf{R})\boldsymbol{r}^{(0)}_j\right), \cdots,\left(\boldsymbol{r}^{(l_{max})}_iD^{(l_{max})}(\mathbf{R})\right)^\top \cdot \left(D^{(l_{max})}(\mathbf{R})\boldsymbol{r}^{(l_{max})}_j\right)\right] \\
 = & ~\left[(r_i^{(0)})^\top r_j^{(0)}, \cdots, (r_i^{(l_{max})})^\top r_j^{(l_{max})}\right] \\
 = & ~ \boldsymbol{r}_i \odot \boldsymbol{r}_j.
\end{align*}

\subsection{Invariance of Message Functions}\label{appen:proofs_msg_func}

Based on the invariance of the above quantities, the invariance of the message functions follows directly from their definitions. Recall Eq.~\eqref{eq:eu_space_msg} and Eq.~\eqref{eq:sh_space_msg}:
\begin{align*}
     m_{EU, ij} := & ~ \phi_{EU} \left( \left[h_i, h_j, \left\|\boldsymbol{x}_{ij} \right\| \right] \right), \\
     m_{SH, ij} := & ~ \phi_{EU} \left( \left[h_i, h_j, \boldsymbol{r}_i \odot \boldsymbol{r}_j \right] \right), \\ 
     m_{EU \to SH, ij} := & ~ \psi_{EU \to SH} ([\boldsymbol{r}_i \odot \boldsymbol{r}_j, \left\|\boldsymbol{x}_{ij} \right\|]]), \\ 
     m_{SH \to E, ij} := & ~ \psi_{SH \to EU} ([\boldsymbol{r}_i \odot \boldsymbol{r}_j, \left\|\boldsymbol{x}_{ij} \right\|]).
\end{align*}

Since $h_i$ and $h_j$ are transformation-invariant scalars, and all the message functions use only invariant quantities ($\|\boldsymbol{x}_{ij}\|$ and $\boldsymbol{r}_i \odot \boldsymbol{r}_j$), we conclude that all the message functions are invariant.

\subsection{Equivariance of the DualEqui Model}\label{appen:proofs_entire_model}

We now prove that the entire DualEqui model is equivariant with respect to E(3) transformations.

\paragraph{Initialization.} In DualEqui, each node $i$ is represented by two types of coordinate vectors: $\boldsymbol{x}_i$ in Euclidean space and $\boldsymbol{r}_i$ in Spherical Harmonics space. The equivariance of $\boldsymbol{x}_i$ is immediate, and, as shown in Eq.~\eqref{eq:eqvar_r_il}, each $\boldsymbol{r}_i^l$ is equivariant. Hence, the initialized node representations are equivariant.

\paragraph{Euclidean Space Update.} In Euclidean space, we update the scalar features $h_i$ and the position vector $\boldsymbol{x}_i$ using updates $\Delta h_{EU,i}$ and $\Delta\boldsymbol{x}_{EU,i}$. The scalar update is given by Eq.~\eqref{eq:delta_h_e}:
\begin{align*}
    \Delta h_{EU, i} := \frac{1}{K} \sum_{k=1}^K \phi_{upd, EU} \left( \left[ h_i, \bigoplus_{j \in \mathcal{N}_{EU}(i)} \alpha_{EU, ij}^k \ \phi_{EU, h}^k (m_{EU, ij}) + \bigoplus_{j \in \mathcal{N}_{SH}(i)} \beta_{EU, ij}^k \psi_{EU, h}^k (m_{SH \to EU, ij}) \right] \right),
\end{align*}
with attention scores defined as:
\begin{align*}
     \alpha_{EU, ij}^k := \sigma\left( \phi_{EU, att}^k (m_{EU, ij}) \right) , 
    \quad 
    \beta_{EU, ij}^k := \sigma\left( \psi_{EU, \beta}^k (m_{SH \to E, ij}) \right).
\end{align*}

Since these computations involve only invariant quantities, $\Delta h_{EU,i}$ is invariant.

Similarly, the position update from Eq.~\eqref{eq:delta_x_e} is:
\begin{align*}
    \Delta \boldsymbol{x}_{EU, i} = \frac{1}{K} \sum_{k=1}^K \left( \bigoplus_{j \in \mathcal{N}_{EU}(i)} \alpha_{EU, ij}^k \ \phi_{EU, x} \left( m_{EU, ij} \right) \boldsymbol{x}_{ij} + \bigoplus_{j \in \mathcal{N}_{SH}(i)} \beta_{EU, ij}^k \psi_{EU, x}(m_{SH \to E, ij}) \boldsymbol{x}_{ij} \right).
\end{align*}

Under an E(3) transformation, the scalar weights are invariant and the aggregation $\oplus$ (i.e., sum or average) is linear, hence:
\begin{align*}
    \Delta \boldsymbol{x}_{EU,i}(\mathbf{R}\boldsymbol{X}+\mathbf{t})=\mathbf{R} \Delta \boldsymbol{x}_{EU,i},
\end{align*}
which proves that $\Delta \boldsymbol{x}_{EU,i}$ is equivariant.

\paragraph{Spherical Harmonics Space Update.} In Spherical Harmonics space, we update the scalar features $h_i$ with $\Delta h_{SH,i}$ and then update the Spherical Harmonics features $\boldsymbol{r}_i$ with $\Delta \boldsymbol{r}_{SH,i}$. The scalar update is defined in Eq.~\eqref{eq:delta_h_sh} as:
\begin{align*}
    \Delta h_{SH, i} := \frac{1}{K} \sum_{k=1}^K \phi_{upd, SH} \left( \left[ h_i, \bigoplus_{j \in \mathcal{N}_{SH}(i)} \alpha_{SH, ij}^k \ \phi_{SH, h}^k (m_{SH, ij}) + \bigoplus_{j \in \mathcal{N}_{EU}(i)} \beta_{SH, ij}^k \psi_{SH, h}^k (m_{EU \to SH, ij}) \right] \right) ,
\end{align*}
where the attention scores are defined as:
\begin{align*}
    \alpha_{SH, ij}^k = \sigma\left( \phi_{SH, att}^k (m_{SH, ij}) \right) , \quad 
\beta_{SH, ij}^k = \sigma\left( \psi_{SH, \beta}^k (m_{EU \to SH, ij}) \right).
\end{align*}

As these updates involve only invariant message functions, $\Delta h_{SH,i}$ is invariant.

For the Spherical Harmonics feature update, we have from Eq.~\eqref{eq:delta_r_sh}:
\begin{align*}
    \Delta \boldsymbol{r}_{SH, i} = \frac{1}{K} \sum_{k=1}^K \left( \bigoplus_{j \in \mathcal{N}_{SH}(i)} \alpha_{SH, ij}^k \phi_{SH, r} (m_{SH, ij}) \boldsymbol{r}_j + \bigoplus_{j \in \mathcal{N}_{EU}(i)} \beta_{SH, ij}^k \psi_{SH, r}(m_{EU \to SH, ij}) \hat{Y}(\boldsymbol{\hat{x}}_{ij}) \right),
\end{align*}
where both $\boldsymbol{r}_j$ and $\hat{Y}(\boldsymbol{x}_{ij})$ are equivariant, transforming under the block diagonal concatenation concatenation of all Wigner-D matrices $[D^{(0)}(\mathbf{R}), D^{(1)}(\mathbf{R}), \cdots, D^{(l_{max})}(\mathbf{R})]$. Therefore,
\begin{align*}
    \Delta \boldsymbol{r}_{SH, i}(\mathbf{R}\boldsymbol{X}+\mathbf{t}) = [D^{(0)}(\mathbf{R}), D^{(1)}(\mathbf{R}), \cdots, D^{(l_{max})}(\mathbf{R})]\Delta \boldsymbol{r}_{SH, i},
\end{align*}
which shows that $\Delta \boldsymbol{r}_{SH, i}$ is equivariant.

\paragraph{Cross-space Interaction Pooling.} In the cross-space interaction pooling stage, we use a biomolecular structure-aware hierarchical pooling strategy to aggregate nodes within a cluster $C$ in both Euclidean space ($\boldsymbol{x}'$) and Spherical Harmonics space ($\boldsymbol{r}'$), along with scalar features $h'$. Since the scalar features are invariant, we focus on the pooling in both vector spaces. 
For the Euclidean space pooling in Eq.~\eqref{eq:x_pooling}, we have:
\begin{align*}
    \boldsymbol{x}'(\mathbf{R}\boldsymbol{X}+\mathbf{t}) = & ~ \frac{1}{|C|} \sum_{i \in C} \alpha_i \left( \mathbf{R}\boldsymbol{x}_i + \gamma \cdot \text{Proj}_{SH}\left(\boldsymbol{r}_i(\mathbf{R}\boldsymbol{X}+\mathbf{t})\right) \right) \\ 
    = & ~ \frac{1}{|C|} \sum_{i \in C} \alpha_i \left( \mathbf{R}\boldsymbol{x}_i + \gamma \cdot  \sum_{l=1}^{L} w_l \| D^{(l)}(\mathbf{R})\boldsymbol{r}_i^l \| \mathbf{R}\boldsymbol{x}_i \right) \\
    = & ~ \frac{1}{|C|} \sum_{i \in C} \alpha_i \left( \mathbf{R}\boldsymbol{x}_i + \gamma \cdot  \sum_{l=1}^{L} w_l \| \boldsymbol{r}_i^l \| \mathbf{R}\boldsymbol{x}_i \right) \\ 
     = & ~ \mathbf{R} \cdot \frac{1}{|C|} \left(\sum_{i \in C} \alpha_i \left( \boldsymbol{x}_i + \gamma \cdot  \sum_{l=1}^{L} w_l \| \boldsymbol{r}_i^l \| \boldsymbol{x}_i \right)\right) \\ 
     = & ~ \mathbf{R} \boldsymbol{x}',
\end{align*}
where $D^{(l)}(\mathbf{R})$ denotes the $l$-order Wigner-D matrix. 

Similarly, for the Spherical Harmonics space pooling in Eq.~\eqref{eq:r_pooling}:
\begin{align*}
    \boldsymbol{r}'(\mathbf{R}\boldsymbol{X}+\mathbf{t}) = & ~ \frac{1}{|C|} \sum_{i \in C} \alpha_i \left( \boldsymbol{r}_i(\mathbf{R}\boldsymbol{X}+\mathbf{t}) + \epsilon \cdot \text{Proj}_{EU}(\mathbf{R}\boldsymbol{x}_i+\mathbf{t}) \right) \\
    = & ~ \frac{1}{|C|} \sum_{i \in C} \alpha_i \left( \boldsymbol{r}_i(\mathbf{R}\boldsymbol{X}+\mathbf{t}) + \epsilon \cdot \hat{Y}( \mathbf{R}(\boldsymbol{x}_i -\boldsymbol{x}_{\text{avg}})) \right) \\
    = & ~ \frac{1}{|C|} \sum_{i \in C} \alpha_i \left( D(\mathbf{R})\boldsymbol{r}_i + \epsilon \cdot D(\mathbf{R})\hat{Y}( \boldsymbol{x}_i -\boldsymbol{x}_{\text{avg}}) \right) \\
    = & ~ D(\mathbf{R}) \cdot \frac{1}{|C|} \left( \sum_{i \in C} \alpha_i \left( \boldsymbol{r}_i + \epsilon \cdot \hat{Y}( \boldsymbol{x}_i -\boldsymbol{x}_{\text{avg}}) \right) \right) \\
    = & ~ D(\mathbf{R}) \boldsymbol{r}',
\end{align*}
where $D(\mathbf{R})$ denotes the block diagonal concatenation of all Wigner-D matrices, i.e., $D(\mathbf{R}):= [D^{(0)}(\mathbf{R}), D^{(1)}(\mathbf{R}), \cdots, D^{(l_{max})}(\mathbf{R})]$. Specifically, $\boldsymbol{x}_{\text{avg}}:=\frac{1}{|C|} \sum_{i \in C} \boldsymbol{x}_i$ in the equation above denotes the center of all input 3D coordinates, and preserves the translation invariance. 

To summarize, we have shown that all node position vectors $\boldsymbol{x}_i$ and $\boldsymbol{r}_i$, and their updates, are equivariant, and that the pooling operations preserve this equivariance. Thus, traversing the entire model structure of DualEqui, we conclude that DualEqui is equivariant with respect to E(3) transformations.
\section{More Experiments and Details}
\label{appen:exp_details}

\subsection{Expressivity on Rotationally Symmetric Structures}
\label{appen:exp_rotsys}
An L-fold symmetric structure is one that remains unchanged under rotation by $360^\circ / L$, exhibiting identical appearance at each of $L$ evenly spaced angles around a central axis. 
Joshi et al.~\cite{joshi2023expressive} demonstrate that spherical tensors of order $L$ cannot discern the orientation of structures with a rotational symmetry greater than $L$-fold.

Table \ref{tab:rotsys} presents the performance of various models on the $L$-fold symmetry classification task for $L = 2, 3, 5$, and $10$. As expected, SchNet, EGNN, and FastEGNN fail to solve the task across all settings due to their lack of higher-order spherical tensors. TFN and HEGNN exhibit correct behavior when equipped with sufficiently high spherical harmonic order (e.g., $l_{\max} \geq 2$), aligning with theoretical expectations from GWL test.

DualEqui achieves similarly expressive results. When $l_{\max} \geq 2$, it consistently attains 100\% accuracy when $L \le l_{max}$, matching the expressive power of TFN/HEGNN.

\begin{table}[h]
\centering
\caption{The expressivity in the $L$-fold task with $L=2, 3, 5, 10$. Anomalous results are highlighted in \colorbox{red!10}{red} and expected in \colorbox{green!10}{green}. DualEqui using the high-order spherical harmonic features is as expressive as TFN and HEGNN.}
\label{tab:rotsys}
\begin{tabular}{lcccc}
\toprule
Folds $L = $         & 2           & 3           & 5           & 10          \\
\midrule
SchNet         & 50.0\footnotesize±0.0  & 50.0\footnotesize±0.0  & 50.0\footnotesize±0.0  & 50.0\footnotesize±0.0  \\
EGNN           & 50.0\footnotesize±0.0  & 50.0\footnotesize±0.0  & 50.0\footnotesize±0.0  & 50.0\footnotesize±0.0  \\
GVPGNN         & 91.5\footnotesize±18.8 & 50.5\footnotesize±5.0  & 66.5\footnotesize±23.5 & 50.0\footnotesize±0.0  \\
FastEGNN       & 50.0\footnotesize±0.0  & 50.0\footnotesize±0.0  & 50.0\footnotesize±0.0  & 50.0\footnotesize±0.0  \\
\midrule
TFN/HEGNN $l_{max}=1$  & \cellcolor{red!10} 50.0\footnotesize±0.0  & 50.0\footnotesize±0.0  & 50.0\footnotesize±0.0  & 50.0\footnotesize±0.0  \\
TFN/HEGNN $l_{max}=2$  & \cellcolor{green!10} 100.0\footnotesize±0.0 & 50.0\footnotesize±0.0  & 50.0\footnotesize±0.0  & 50.0\footnotesize±0.0  \\
TFN/HEGNN $l_{max}=3$  & \cellcolor{green!10} 100.0\footnotesize±0.0 & \cellcolor{green!10} 100.0\footnotesize±0.0 & 50.0\footnotesize±0.0  & 50.0\footnotesize±0.0  \\
TFN/HEGNN $l_{max}=5$  & \cellcolor{green!10} 100.0\footnotesize±0.0 & \cellcolor{green!10} 100.0\footnotesize±0.0 & \cellcolor{green!10} 100.0\footnotesize±0.0 & 50.0\footnotesize±0.0  \\
TFN/HEGNN $l_{max}=10$ & \cellcolor{green!10} 100.0\footnotesize±0.0 & \cellcolor{green!10} 100.0\footnotesize±0.0 & \cellcolor{green!10} 100.0\footnotesize±0.0 & \cellcolor{green!10} 100.0\footnotesize±0.0 \\
\midrule
DualEqui $l_{max}=1$   & \cellcolor{red!10} 50.5\footnotesize±5.0  & 50.0\footnotesize±0.0  & 50.0\footnotesize±0.0  & 50.5\footnotesize±5.0  \\
DualEqui $l_{max}=2$   & \cellcolor{green!10} 100.0\footnotesize±0.0 & 50.0\footnotesize±0.0  & 63.5\footnotesize±22.2 & 95.5\footnotesize±14.3 \\
DualEqui $l_{max}=3$   & \cellcolor{green!10} 100.0\footnotesize±0.0 & \cellcolor{green!10} 100.0\footnotesize±0.0 & 55.0\footnotesize±15.0 & 60.0\footnotesize±20.0 \\
DualEqui $l_{max}=5$   & \cellcolor{green!10} 100.0\footnotesize±0.0 & \cellcolor{green!10} 100.0\footnotesize±0.0 & \cellcolor{green!10} 100.0\footnotesize±0.0 & 50.0\footnotesize±0.0  \\
DualEqui $l_{max}=10$  & \cellcolor{green!10} 100.0\footnotesize±0.0 & \cellcolor{green!10} 100.0\footnotesize±0.0 & \cellcolor{green!10} 100.0\footnotesize±0.0 & \cellcolor{green!10} 100.0\footnotesize±0.0  \\
\bottomrule
\end{tabular}
\end{table}

\subsection{Computational Efficiency Analysis}

To assess the computational efficiency of DualEquiNet, we compare the training time per epoch across several baseline models, as shown in Table~\ref{tab:my-table}. While DualEquiNet introduces additional overhead due to its dual-space message passing and spherical harmonics computations, it remains significantly more efficient than high-order tensor-based models like TFN, with training time comparable to FastEGNN and GotenNet. This demonstrates that our design strikes a favorable balance between computational cost and model expressivity, enabling scalable training on large biomolecular systems.

\begin{table}[h]
\caption{Training time of each model (seconds/epoch). }
\label{tab:my-table}
\centering
\begin{tabular}{cc}
\toprule
Model    & Train Time (s/epoch) \\
\midrule 
SchNet   & 4.361                \\
EGNN     & 6.240                \\
GVPGNN   & 11.172               \\
TFN      & 56.235               \\
FastEGNN & 11.755               \\
HEGNN    & 6.320                \\
GotenNet & 10.332               \\
DualEqui & 10.990               \\
\bottomrule
\end{tabular}%
\end{table}

\subsection{Datasets Details}
\label{appen:datasets_detials}

\paragraph{COVID Vaccine} 
The OpenVaccine COVID-19 dataset~\cite{wayment2022deep} consists of 4,082 synthetic RNA sequences, each 107-130 nucleotides long. It was developed to support the predictive modeling of RNA stability for mRNA vaccine development. Each nucleotide is annotated with three quantitative properties: reactivity, degradation under pH10, and degradation under Mg pH10 conditions. These measurements reflect how susceptible each nucleotide is to chemical degradation and serve as surrogates for RNA structural stability. Sequences with low signal-to-noise ratio (SNR < 1) are filtered out to ensure label reliability, following prior best practices. This dataset is especially suited for testing fine-grained, residue-level predictions. In our experiments, we infer the 3D atomic structure for each sequence using RhoFold~\cite{shen2022e2efold}.

\paragraph{Ribonanza} 
The Ribonanza-2k dataset~\cite{he2024ribonanza} includes 2,260 RNA sequences, each around 177 nucleotides in length, annotated at the nucleotide level with reactivity values obtained from two chemical probing agents: DMS and 2A3. These chemical modifications reflect the accessibility and flexibility of each nucleotide, thereby encoding aspects of RNA's secondary and tertiary structures. The labels thus capture nuanced differences in base-pairing and folding landscapes across diverse RNA sequences, making the dataset a benchmark for modeling structural flexibility.

\paragraph{Tc-Riboswitches} 
The Tc-Riboswitches dataset~\cite{groher2018tuning} consists of 355 mRNA sequences, each 67-73 nucleotides long, paired with graph-level regression labels reflecting gene regulatory behavior in response to tetracycline binding. Specifically, the label represents the expression shift induced by the presence of a tetracycline switching factor, an RNA-based synthetic regulatory system widely studied in gene therapy and synthetic biology. Unlike COVIDVaccine and Ribonanza, which have residue-level labels, Tcribo focuses on sequence-level prediction, allowing evaluation of how local structure aggregates into global function.

\paragraph{RNASolo SASA}
The SASA dataset consists of RNA structures annotated with Solvent-Accessible Surface Area (SASA) values for each nucleotide. SASA quantifies the extent to which each nucleotide is exposed to solvent, providing important insights into RNA folding, stability, and interaction potential. We compute per-nucleotide SASA values by aggregating atomic-level SASA from experimentally resolved PDB structures in the RNASolo database~\cite{adamczyk2022rnasolo}. These values are extracted using DSSR, which internally implements the Shrake and Rupley algorithm~\cite{shrake1973environment}. This method models solvent accessibility by rolling a spherical probe over the van der Waals surface of each atom, and then estimating the exposed area based on point sampling. The resulting SASA values reflect the 3D packing and surface topology of RNA molecules. Unlike earlier tasks focused on sequence-derived stability, RNASolo-SASA emphasizes the role of 3D structural precision and surface geometry. The dataset includes a diverse set of RNA structures with variable lengths and topologies.

\paragraph{mRFP Protein SASA}
We utilize the mRFP protein dataset introduced by Stanton et al.\cite{stanton2022accelerating} for the task of SASA prediction. This dataset was designed as part of a realistic large-molecule sequence design benchmark, aiming to simulate practical protein engineering objectives, consisting of approximately 200-residue red fluorescent proteins (RFPs), where each protein sequence is annotated with ground-truth SASA values. We randomly select 718 proteins and the ground-truth 3D structures are predicted by AlphaFold2~\cite{jumper2021highly}.

\paragraph{TorsionAngle}
\label{appen:datasets_detials_torsion}
The TorsionAngle dataset is designed to assess a model's ability to recover fine-grained structural geometry by predicting residue-level torsion angles, including backbone and glycosidic dihedrals. These angles—such as $\alpha, \beta, \gamma, \delta, \epsilon, \zeta$, and $\chi$—are crucial descriptors of RNA conformation and govern higher-order folding. We extract these angles from RNA structures in the RNASolo database also using the DSSR, which computes torsions directly from atomic coordinates. Each sample in the dataset contains the full 3D atomic structure and a set of seven torsion angles per nucleotide. Because RNA torsions are sensitive to both local backbone rigidity and global conformation, this dataset provides a challenging test for models to capture both short-range and long-range structural dependencies. All angles are normalized into the range $[-180^\circ, 180^\circ]$, and training uses a circular MAE loss described in Section~\ref{sec:exp_torsionangle}. 

Here, we summarize all the datasets in Table~\ref{tab:dataset_stat}.

\begin{table}[h]
\caption{Dataset Statistics.}
\label{tab:dataset_stat}
\resizebox{\textwidth}{!}{
\begin{tabular}{ccccccc}
\toprule
                                                                    & \textbf{COVID}                                                       & \textbf{Ribonanza} & \textbf{Tc-Riboswitches} & \textbf{RNASolo SASA}  & \textbf{mRFP Protein SASA}           & \textbf{TorsionAngle}     \\
\midrule
Task Level                                                          & residue-level                                                        & residue-level      & graph-level              & residue-level      & graph-level           & residue-level \\
Target                                                              & \begin{tabular}[c]{@{}c@{}}reactivity \& \\ degradation\end{tabular} & reactivity         & switching factor         & SASA Value         & SASA Value            & torsion angles   \\
\# Sequences                                                        & 4082                                                                 & 2260               & 355                      & 1018               & 718                   & 1018             \\
Sequence Length                                                     & 107 - 130                                                            & 177                & 66 - 75                  & 10 - 400           & 216 - 217             & 10 - 400         \\
\# Labels                                                           & 3                                                                    & 2                  & 1                        & 1                  & 1                     & 7                \\
\begin{tabular}[c]{@{}c@{}}\# Avg. Atoms \end{tabular}              & 363                                                                  & 531                & 216                      & 249                & 649                   & 249              \\
\bottomrule
\end{tabular}
}
\end{table}

\subsection{Comparison of Baselines}
\label{appen:compare_baselines}

\textbf{SchNet} is an invariant GNN using continuous-filter convolutional layers based on pairwise distances expanded in radial bases. It employs only scalar features, making predictions rotation- and translation-invariant. It lacks higher-order directional features and handles long-range interactions primarily through stacking layers or increasing cutoff distances.

\textbf{EGNN} achieves E(3)-equivariance by scalarizing messages computed from pairwise distances and features to scale relative position vectors, updating Cartesian coordinates. Unlike SchNet, it explicitly uses first-order geometric (Cartesian) features, capturing directional information. Long-range interactions are possible via dense connections, though it can become computationally expensive for large systems.

\textbf{GVPGNN} maintains separate scalar and Cartesian vector features, updating them with an equivariant vector perceptron that couples vector updates with scalar invariants. This ensures SE(3)-equivariance and improves geometric expressiveness over EGNN. It remains constrained to first-order Cartesian vectors and, like EGNN, relies on standard message passing for long-range information.

\textbf{TFN} utilizes spherical harmonics and Clebsch–Gordan tensor products to explicitly represent higher-order geometric correlations, achieving rigorous SE(3)-equivariance. While highly expressive, TFN’s complexity is significantly greater than scalar-based methods. It primarily captures local interactions, with long-range interactions requiring deep stacking or additional global modules.

\textbf{FastEGNN} extends EGNN by incorporating virtual nodes that aggregate and distribute global information, enhancing long-range communication without substantially increasing computational complexity. It preserves EGNN’s E(3)-equivariance and Cartesian features, improving efficiency on larger or more distant interactions while maintaining EGNN’s directional limitations.

\textbf{HEGNN} enhances EGNN’s framework by integrating higher-order spherical harmonic features (beyond first-order vectors). It scalarizes tensor interactions for computational efficiency, achieving improved geometric expressiveness. While computationally manageable, it remains primarily focused on local interactions unless combined with specialized long-range mechanisms.

\textbf{GotenNet} employs geometry-aware attention with learned higher-order Cartesian tensors, avoiding explicit spherical harmonic computations while maintaining E(3)-equivariance. Its attention mechanism efficiently captures long-range interactions and high-order geometric patterns, combining the benefits of expressive tensors with scalable global communication.

Table~\ref{tab:baseline_comparison} summarizes key properties of baseline models in terms of equivariance, feature representation, spherical harmonic usage, and ability to model long-range interactions. While many recent models achieve E(3)-equivariance, they differ in expressivity and scalability. Invariant models like SchNet rely solely on scalar features and lack the capacity to model directional or long-range dependencies. Scalarization-based equivariant models such as EGNN and GVP-GNN use Cartesian vectors but are limited to local neighborhoods. High-order methods like TFN offer expressive SH-based tensors but suffer from inefficient tensor product calculations. In contrast, HEGNN, GotenNet, and DualEqui provide efficient SH updates, with DualEqui and FastEGNN uniquely supporting long-range modeling. DualEqui stands out for combining spherical representations with long-range communication, enabled by its cross-space design.

\begin{table}[h]
\centering
\caption{Comparison of baseline geometric GNN models. ``Long-Range” indicates whether the model has a built-in mechanism to capture long-range interactions efficiently. Designs with higher expressivity or efficiency are highlighted in  \colorbox{green!10}{green}.}
\label{tab:baseline_comparison}
\begin{tabular}{lcccc}
\toprule
\textbf{Model} & \textbf{Equivariance} & \textbf{Feature Type} & \textbf{Efficient SH Update} & \textbf{Long-Range} \\
\midrule
\textbf{SchNet}     & Invariant    & Scalar         &               & $\times$  \\
\textbf{EGNN}       & \cellcolor{green!10} Equivariant  & Cartesian      &               & $\times$   \\
\textbf{GVP-GNN}    & \cellcolor{green!10} Equivariant  & Cartesian      &               & $\times$  \\
\textbf{TFN}        & \cellcolor{green!10} Equivariant  & \cellcolor{green!10} Spherical      &  $\times$     & $\times$  \\
\textbf{FastEGNN}   & \cellcolor{green!10} Equivariant  & Cartesian      &               & \cellcolor{green!10} \checkmark  \\
\textbf{HEGNN}      & \cellcolor{green!10} Equivariant  & \cellcolor{green!10} Spherical      & \cellcolor{green!10} \checkmark    & $\times$  \\
\textbf{GotenNet}   & \cellcolor{green!10} Equivariant  & \cellcolor{green!10} Spherical      & \cellcolor{green!10} \checkmark    & $\times$ \\
\textbf{DualEqui}   & \cellcolor{green!10} Equivariant  & \cellcolor{green!10} Spherical      & \cellcolor{green!10} \checkmark    & \cellcolor{green!10} \checkmark \\
\bottomrule
\end{tabular}
\end{table}

\subsection{Reproducibility}
All experiments in this work are conducted under unified and controlled conditions to ensure fair comparison and reproducibility. For each dataset, we adopt an 8:1:1 split for training, validation, and testing. All reported results are averaged over 5 independent random splits, with standard deviations included. For each run, the best-performing model is selected based on validation set performance and evaluated on the test set. Our implementation is built on PyTorch and PyTorch Geometric~\cite{Fey/Lenssen/2019}. All code and datasets will be publicly available upon acceptance.

We apply consistent hyperparameter tuning for all models—including baselines and our DualEqui—using the Optuna~\cite{akiba2019optuna} framework with 100 trials per method. Search spaces and best configurations are detailed in Appendix D.5. To ensure consistency, we fix random seeds for all runs (PyTorch, NumPy, CUDA) and repeat each experiment across 5 different seeds. All experiments were conducted on a single NVIDIA A6000 GPU with a number of parameters limit up to $4 \times 10^6$. In Table~\ref{tab:hyperparameters}, we provide our searched optimal hyperparameters of DualEqui on each dataset. More detialed training scripts will be released with the code.

\begin{table}[]
\caption{Optimal hyperparameters of DualEqui on each dataset.}
\label{tab:hyperparameters}
\centering
\resizebox{\textwidth}{!}{
\begin{tabular}{ccccccc}
\hline
\textbf{Dataset}    & \textbf{Covid} & \textbf{Ribonanza} & \textbf{Tcribo} & \textbf{RNASolo SASA} & \textbf{mRFP Protein SASA} & \textbf{TorsionAngle} \\ \hline
$L_{atom}$          & 4              & 1                  & 3               & 4                     & 1                          & 3                     \\
$L_{nt}$            & 3              & 4                  & 3               & 1                     & 1                          & 4                     \\
\# heads            & 2              & 8                  & 4               & 6                     & 3                          & 5                     \\
hidden              & 72             & 128                & 84              & 150                   & 174                        & 140                   \\
atom $d_{EU}$       & 4.82           & 3.93               & 3.03            & 10.44                 & 3.68                       & 8.93                  \\
residue $d_{EU}$    & 26.4           & 30.61              & 122.64          & 47.02                 & 116.39                     & 42.45                 \\
lr                  & 0.0030         & 0.0015             & 0.0002          & 0.0002                & 0.0005                     & 0.0013                \\
weight decay        & 1.37e-6        & 5.65e-8            & 1.17e-7         & 1.37e-5               & 3.00e-7                    & 2.10e-7               \\
$l_{max}$           & 2              & 2                  & 2               & 2                     & 3                          & 3                     \\
atom $d_{SH}$       & 0.97           & 1.0                & 0.95            & 1.00                  & 0.96                       & 0.99                  \\
residue $d_{SH}$    & 0.97           & 0.95               & 0.92            & 0.95                  & 0.95                       & 0.91                  \\
batch size          & 32             & 25                 & 32              & 16                    & 16                         & 16                    \\ 
epochs              & 1000           & 1000               & 1000            & 1000                  & 1000                       & 1000                  \\ 
\hline
\end{tabular}
}
\end{table}

\end{document}